\documentclass[twocolumn]{aastex7}
\usepackage{natbib}
\usepackage{booktabs}
\usepackage{float}

\usepackage{amsmath}

\newcommand{\othree}{[O~{\sc iii}]}

\newcommand{\cfour}{C~{\sc iv}}
\newcommand{\sifour}{Si~{\sc iv}}
\newcommand{\hb}{H$\beta$}

\newcommand{\kms}{km s$^{-1}$}
\newcommand{\alphao}{$\alpha_{\rm o}$}
\newcommand{\alphaoi}{$\alpha_{oi}$}
\newcommand{\alphair}{$\alpha_{ir}$}

\shorttitle{SEDs of BAL Quasars at High Redshift}
\shortauthors{Ahmed et al.}

\begin{document}

\title{Exploring the Spectral Energy Distributions of Luminous Broad Absorption Line Quasars at High Redshift}

\author[0009-0008-0969-4084]{Harum Ahmed}
\affiliation {Department of Physics, University of North Texas, Denton, TX 76203, USA} \email{HarumAhmed@my.unt.edu}

\author[0000-0001-6217-8101]{Sarah C. Gallagher}
\affiliation{Department of Physics \& Astronomy, University of Western Ontario, 1151 Richmond St, London, ON N6C 1T7, Canada} \email{sgalla4@uwo.ca} 
\affiliation{Institute for Earth and Space Exploration, The University of Western Ontario, London, ON, N6A 3K7, Canada}

\author[0000-0003-4327-1460]{Ohad Shemmer}
\email{ohad@unt.edu} 
\affiliation{Department of Physics, University of North Texas, Denton, TX 76203, USA}

\author[0000-0002-1207-0909]{Michael S. Brotherton}
\email{mbrother@uwyo.edu} 
\affiliation{Department of Physics and Astronomy, University of Wyoming, Laramie, WY 82071, USA}

\author[0000-0003-0192-1840]{Cooper Dix}
\email{cooper.dix@austin.utexas.edu} 
\affiliation{Department of Physics, University of North Texas, Denton, TX 76203, USA}
\affiliation{Department of Astronomy, The University of Texas at Austin, Austin, TX 78712, USA}

\author[0009-0007-4337-4239]{Leigh Parrott}
\email{LeighParrott@my.unt.edu} 
\affiliation{Department of Physics, University of North Texas, Denton, TX 76203, USA}

\author[0000-0002-1061-1804]{Gordon T. Richards}
\email{gtr@physics.drexel.edu} 
\affiliation{Department of Physics, Drexel University, 32 S. 32nd Street, Philadelphia, PA 19104, USA}

\received{2025 January 02}
\revised{2025 March 27}
\accepted{2025 April 03}

\begin{abstract}

We present the multiwavelength spectral energy distributions (SEDs) for 65 luminous broad absorption line (BAL) quasars with redshifts \hbox{$1.55 \lesssim z \lesssim 3.50$} from the Gemini Near Infrared Spectrograph - Distant Quasar Survey (GNIRS-DQS). We integrate data from a variety of ground- and space-based observatories to construct a comprehensive spectral profile of these objects from radio through X-rays. In addition, we present a mid-infrared to X-ray composite SED of these sources. Our dataset represents the most uniform sample of BAL quasars, providing a statistically robust set of SEDs. Our findings indicate that the BAL quasars in the GNIRS-DQS sample exhibit significant reddening in the ultraviolet-optical continuum relative to their non-BAL counterparts, consistent with previous studies. Notably, our analysis reveals no significant differences in the mid- or near-infrared spectral regime between BAL and non-BAL quasars. In line with previous work, we find no strong evidence that BAL and non-BAL quasars possess fundamentally different SEDs, also consistent with recent findings that both groups display similar rest-frame optical emission-line properties.

\end{abstract}


\section{Introduction} \label{sec:intro}

Broad Absorption Line (BAL) quasars, which are identified in approximately 10-25\% of sources in optically selected quasar samples (e.g., \citealt{Reichard03}; \citealt{Hewett03}; \citealt{Trump06}; \citealt{Ganguly08}; \citealt{Knigge08}; \citealt{Ahmed24}), exhibit pronounced, blueshifted absorption features in high-ionization emission lines such as \cfour~$\lambda 1549$, \sifour~$\lambda1393$, and N~{\sc v}~$\lambda1240$. These features arise from outflowing material along the line of sight, with velocities ranging from a few thousand \hbox{\kms}~to over 50,000 \kms~(e.g., \citealt{Wey91}; \citealt{Hall02}; \citealt{Bruni19}; \citealt{Rodriguez20}; \citealt{Bischetti23}). The strength of these absorption features are characterized by metrics such as the ``Balnicity Index" (BI) or ``Absorption Index" (AI), which requires a minimum velocity width of \hbox{2000 \kms}~or 450 \kms~ at 10\% depth below the UV continuum, respectively (e.g., \citealt{Wey91}; \citealt{Hall02}).

These quasar-driven outflows, which are essential components of galaxy formation models (e.g., \citealt{Silk98}), play a key role in active galactic nucleus (AGN)-driven feedback by regulating star formation, redistributing interstellar gas, and influencing the co-evolution of quasars and their host galaxies (e.g., \citealt{Hopkins05}; \citealt{Hamann13}; \citealt{Begelman06}). Such outflows make BAL quasars crucial for understanding AGN feedback mechanisms in galaxy evolution.

To comprehensively understand the mechanisms driving these outflows and their broader implications, it is crucial to investigate the spectral energy distribution (SED) of BAL quasars. Quasar SEDs span the entire electromagnetic spectrum, from hard X-rays to the radio band (e.g., \citealt{Elvis94}; \citealt{Richards06}; \citealt[hereafter K13]{K13}),  and reveal multiple emission components. These components include the accretion disk emission, which is characterized by a blue ultraviolet (UV)-optical continuum spectrum; the torus, emitting reprocessed continuum in the infrared (IR) with a break at 1 $\mu$m; and X-ray emission, which itself comprises at least two components: the soft excess and the Compton ``hump." The mechanisms driving these X-ray features remain under debate, with models invoking reflection off the inner disk or intrinsic flux from the corona (e.g., \citealt{Giustini19}). The SED conveys the essential physical properties of the accreting black hole system, such as the mass, spin, and accretion rate influenced by orientation. Untangling the relationship between these physical properties and their spectral signatures is the key objective of quasar SED studies.

A composite SED, spanning wavelengths from radio to X-rays, of a sample of 109 Palomar Green (PG) quasars was presented in \citet[see also \citealt{Elvis94}]{Sanders89}. Subsequent studies have expanded the number of quasar SEDs analyzed (e.g., \citealt{Shang11}; \citealt{Bianchini19}). Notably, \citet{Richards06} computed the mean SED from a sample of 259 quasars with both Sloan Digital Sky Survey (SDSS; \citealt{York00}) and Spitzer Infrared Array Camera photometry. K13 extended their work using a sample of 119,652 non-reddened type 1 quasars at $0.064 < z < 5.46$ with at least SDSS photometry, constructing a mean SED between 10 keV and $\sim30~\mu$m, making it the most robust due to the number of sources used. Both \citet{Richards06} and K13 constructed composite SEDs for sample subsets based on quasar luminosity. \citet{Richards06} note subtle differences indicating luminosity affects the overall SED shapes, particularly in the mid-IR. Similarly, K13 reveals that, compared to SEDs of high-luminosity sources, SEDs of low-luminosity sources display a harder (bluer) far-UV slope, a redder optical continuum, and reduced hot dust emission.

The majority of studies on SEDs in quasars focus on unabsorbed (non-BAL) quasars due to the increased evidence of dust reddening and extinction in the UV-optical spectra of BAL quasars. Additionally, most BAL quasars exhibit low observed X-ray fluxes compared to their predicted values based on their optical fluxes, making it challenging to create a comprehensive multiwavelength representation of these sources. However, a notable study by \citet[hereafter G07]{Gall07a} presented SEDs for a sample of 38 BAL quasars from the Large Bright Quasar Survey (LBQS) at $1.5 \lesssim z \lesssim 2.9$, covering radio to \hbox{X-ray} wavelengths. In their work, they find no substantial evidence for inherent differences between the SEDs of BAL and non-BAL quasars with comparable luminosities.

Investigating SEDs in high-luminosity quasars is crucial, as both theoretical models and observational evidence indicate that these sources exhibit the most powerful feedback mechanisms (e.g., \citealt{Fiore17}; \citealt{Bischetti19}). In this work, we focus on 65 high-luminosity, high-redshift BAL quasars from the Gemini Near Infrared Spectrograph - Distant Quasar Survey (GNIRS-DQS) (e.g., \citealt{M21}; \citealt[hereafter M23]{M23}), offering extensive multiwavelength coverage. This investigation offers the most comprehensive and uniformly constructed dataset of SEDs for BAL quasars, providing a critical resource for understanding the BAL phenomenon.

This paper is organized as follows: Section \ref{sec:data} presents the multiwavelength data used in the construction of the SEDs. Section \ref{sec:results} presents the SEDs, the composites, and the continuum properties. Section \ref{sec:conclusion} gives a brief summary of results and conclusions. Throughout this work, we adopt a $\Lambda$CDM cosmology with \hbox{$H_{0}=70$ km s$^{-1}$ Mpc$^{-1}$}, $\Omega_{\rm M}=0.3$, and $\Omega_{\Lambda}=0.7$ when calculating quantities such as quasar luminosities or luminosity distances (e.g., \citealt{Spergel07}). 

\section{Multiwavelength Data} \label{sec:data}

We selected sources from the GNIRS-DQS catalog\footnote{https://datalab.noirlab.edu/gnirs\textunderscore dqs.php} which constitutes the largest, uniform inventory of rest-frame optical spectral properties of luminous quasars at high redshift. Specifically, the catalog consists of 260 SDSS quasars, including 65 BAL and 195 non-BAL quasars, spanning a redshift range of \hbox{$1.55 \lesssim z \lesssim 3.50$}, with \hbox{$-28.0 \lesssim M_i \lesssim -30.0$ mag} (M23). The rest-frame monochromatic luminosities ($\nu L_{\nu}$) for these sources range from \hbox{$\sim 10^{46} - 10^{47}$ erg~s$^{-1}$}, based on continuum flux density measurements at rest-frame $5100~\rm \AA$ (M23). By ensuring that the BAL and non-BAL quasars are matched in $\nu L_{5100 \rm \AA}$ and are sourced from the same survey, we effectively mitigate potential selection biases. 

The sample of 65 quasars predominantly comprises of high-ionization BAL (HiBAL) quasars, with four low-ionization BAL quasars (LoBAL), which exhibit additional absorption in low-ionization emission lines such as Mg~{\sc ii} $\lambda$2803, and Al~{\sc iii} $\lambda$1857, as identified in \citet{Ahmed24}. The BAL quasar properties are given in Table \ref{tab:table1}.

\startlongtable
\begin{deluxetable*}{lccccccccccccccc}
\tablenum{1}
\tablecaption{GNIRS-DQS BAL Quasar Sample Properties \label{tab:table1}}
\tablecolumns{11}
\setlength{\tabcolsep}{2.8 pt}
\tablewidth{0.5 cm}
\tabletypesize{\scriptsize}
\tablehead{
\colhead{Quasar} & \colhead{}  & \colhead{} & \colhead{} & \colhead{} & \colhead{} &  \multicolumn{4}{r}{log $L_{\nu}$(erg $\rm s^{-1}~Hz^{-1}$)} & \colhead{}\\
\cmidrule(lr){8-10}
 \colhead{(SDSS J)} & \colhead{${z_{\rm sys}}^{a}$}  & \colhead{${E(B - V)}^{b}$}  & \colhead{$R^{c}$} & \colhead{${\alpha_{\rm o}^{d}}$} & \colhead{{\alphaoi}$^{e}$} & \colhead{{\alphair}$^{f}$}  & \colhead{2500 \AA} & \colhead{5100 \AA} & \colhead{3 $\mu$m} & \colhead{log $L_{\rm bol}$ {$(\rm erg~s^{-1})$$^{g}$}}\
}

\startdata
$001249.89+285552.6$ & 3.233 & 0.042 & \nodata & -1.25 & -1.73 & -1.02  & 30.61 & 31.38 & 32.09 & 46.98\\
$001355.10-012304.0$ & 3.380 & 0.059 & 0.67 & -0.60 & -2.04 & -1.79  & 30.41 & 31.18 & 32.11 & 47.07\\
$004613.54+010425.7$ & 2.165 & 0.071 & 2.64 & -0.88 & -1.17 & -2.28  & 30.44 & 31.21 & 31.61 & 46.87\\
$013012.36+153157.9$ & 2.343 & 0.050 & 12.71 & -0.96 & -1.65 & -0.70 & 30.28 & 31.06 & 31.80 & 46.81\\
$013652.52+122501.5$ & 2.388 & 0.083 & 1.41 & -0.78 & -2.12 & -1.13  & 30.39 & 31.17 & 32.26 & 46.92\\
\enddata
\tablecomments{This table is available in its entirety in machine-readable format.}
\tablenotetext{a}{Systemic redshift from M23 (Table 2, Column 2).}
\tablenotetext{b}{Intrinsic extinction values calculated using the Small Magellanic Cloud (SMC) law (see Section \ref{subsec:comp}).}
\tablenotetext{c}{Radio-loudness parameter, $R = f_{\nu, 5\rm GHz} / f_{\nu, \rm 4400\AA}$ calculated using radio data (see Section \ref{sec:data}).}
\tablenotetext{d}{Optical power-law index ($L_{\nu}\propto \nu^{\alpha_{\rm o}}$) from a fit to the rest-frame 1200 and 5000 \AA\ photometry (see Section \ref{subsec:alphao}).}
\tablenotetext{e}{Point-to-point flux density slope between 5100 \AA\ and 3 $\mu$m flux density (see Section \ref{subsec:alphaoi}).}
\tablenotetext{f}{Mid-IR power-law index ($L_{\nu}\propto \nu^{\alpha_{ir}}$) from a fit to the rest-frame 1.8-10 $\mu$m photometry (see Section \ref{subsec:alphair}).}
\tablenotetext{g}{Bolometric luminosity integrated from 1 $\mu$m to 2 keV (see Section \ref{subsec:BC}).}
\end{deluxetable*}
\raggedbottom

This multiwavelength SED work comprises mid-and near-IR photometry obtained from the Wide-field Infrared Survey Explorer (WISE; e.g., \citealt{Wright10}) and the Two Micron All Sky Survey (2MASS; \citealt{Skrutskie06}) for all 65 BAL quasars. These datasets have been augmented with multiband optical photometry for all sources, nearly complete radio observations from the Very Large Array (VLA; 64/65 sources), rest-frame UV data (54/65 sources), and X-ray data available for 23 of the BAL sources.

Additionally, for the 195 non-BAL quasars, we follow the same methodology as BAL quasars to obtain WISE, 2MASS, and SDSS photometry, along with UV data for 125 sources and X-ray data for 63 sources (A. Marlar, priv. comm.). The GNIRS-DQS non-BAL quasars have more extensive X-ray coverage than BAL sources, as the latter are known to be weak X-ray emitters (e.g., \citealt{Gall02}, \citeyear{Gall06}; \citealt{Wang22}). We exclude radio data for non-BAL quasars, as our analysis does not primarily focus on radio properties.

All data were compiled from publicly available archives. This section provides a brief overview of the data sources and acquisition methods for the SED data.

\subsection{Radio}

Radio data for this work are sourced from the Faint Images of the Radio Sky at Twenty cm (FIRST\footnote{http://www.sundog.stsci.edu}; \citealt{Becker95}; \citealt{White97}) and the National Radio Astronomical Observatory VLA Sky Survey (NVSS\footnote{http://www.cv.nrao.edu/nvss}; \citealt{Condon98}) both at 1.4 GHz. These data are publicly available for 64 out of 65 BAL quasars and are presented in Table \ref{tab:table2}. Upper limits for non-detections (at $5\sigma$) are set at 1 mJy in the FIRST survey and 2.5 mJy in the NVSS survey (\citealt{Condon98}). Given its higher resolution and sensitivity, we prioritize FIRST measurements where available, using NVSS data only for targets lacking FIRST coverage.

One source, SDSS J001249.89+285552.6, lacks radio coverage within $\sim30''$ of the target coordinates in the FIRST and NVSS footprints. Additionally, all sources were cross-checked with the NASA/IPAC Extragalactic Database (NED) to confirm the corresponding radio data. Note that one BAL quasar in our sample, SDSS J114705.24+083900.6, is also classified as a radio-loud (RL)\footnote{We consider radio-loud quasars to have $R > 100$, where $R$ is the ratio of the flux densities at 5 GHz and 4400 \AA; \citealt{Kellermann89}).} quasar based on radio data from NVSS (see Table \ref{tab:table1}).

\subsection{Mid-Infrared}

We extended our SEDs into the mid-IR using data from the WISE final data release\footnote{https://wise2.ipac.caltech.edu} (\citealt{Wright10}). The WISE bandpasses, designated as W1 through W4, correspond to effective observed frame wavelengths of 3.4, 4.6, 12.0, and 22.0 $\mu$m, respectively, which are pertinent for characterizing quasar SEDs. Since \citet{Lyke20} only includes W1 and W2 bandpasses for these sources, we performed matching with the WISE final data release by identifying all WISE point sources within $2''$ of an SDSS quasar. This matching radius maximizes the number of true objects matched while also minimizing the number of false matches. All quasars were detected in all four WISE bands and the flux densities for all 65 sources are given in Table \ref{tab:table2}.

\subsection{Near-Infrared}

By selection, all GNIRS-DQS sources have a 2MASS detection in at least one band, $J$, $H$, or $K$, corresponding to central wavelengths of 1.25 $\mu$m, 1.65 $\mu$m, or \hbox{2.16 $\mu$m}, respectively. Therefore, all 65 BAL quasars are included in the 2MASS\footnote{http://www.ipac.caltech.edu/2mass} point source catalogs and in \citet{Lyke20}. These include values that were matched to the 2MASS All-Sky source catalog using a matching radius of $2''$. All 65 BAL quasars were detected in all three bands and the data are listed in Table \ref{tab:table2}.

\subsection{Optical}

Since the GNIRS-DQS sources are selected from \citet{Lyke20}, all 65 BAL sources have available SDSS photometry in the five optical bandpasses \textit{ugriz} (\citealt{Fukugita96}). We present the \textit{ugriz} Point Spread Function (PSF) flux densities for these objects in Table \ref{tab:table3}, converted to units of mJy using the AB magnitude system \citep{Oke83}, in which the zero-point flux density is standardized to 3631 Jy for all bandpasses\footnote{We note that small offsets relative to the AB system exist in the \textit{ugriz} bands due to calibration practices (e.g., \citealt{Padmanabhan08}; \citealt{Doi10}). For consistency, our conversions assume the nominal zero-point but should be interpreted with this caveat in mind.}. All the flux densities have been corrected for Galactic extinction according to \citet{Schlegel98} with corrections to the extinction coefficients as given by \citet{Schlafly11}.

\subsection{Ultraviolet}

To incorporate UV flux densities into the SEDs, we utilize data from the Galaxy Evolution Explorer (GALEX\footnote{https://galex.stsci.edu/GR6/}; \citealt{Martin05}) when available. The effective wavelengths for the near-UV (NUV) and far-UV (FUV) bandpasses are 2267 \AA\ and 1516 \AA, respectively. Among the 65 BAL quasars, our matched sample with GALEX includes 54 sources. Specifically, 12 sources have both FUV and NUV photometry, 39 sources have data only in the NUV band, and three sources have data exclusively in the FUV band. All GALEX photometry has been corrected for Galactic extinction, assuming $A_{\rm NUV} = -8.741 \times E(B - V)$ and $A_{\rm FUV} = 8.24 \times E(B - V) - 0.67 \times E(B - V)^2$ (e.g., \citealt{Wyder07}) and are listed in Table \ref{tab:table3}.

\subsection{X-ray}

Because BAL quasars are known to be weak X-ray emitters, the number of such sources with X-ray detections in large surveys is quite small compared to the size of the respective samples in the optical and IR. For example, only four of the 65 BAL sources were targeted and detected with Chandra X-ray Observatory and are available in the Chandra Point Source Catalog\footnote{http://cda.cfa.harvard.edu/cscview/} (\citealt{Evans10}) in the 0.2-2 keV band. Additionally, the 4XMM XMM-Newton Serendipitous Source Catalog\footnote{http://xmmssc.irap.omp.eu/Catalogue/4XMM-DR14/} (\citealt{Webb20}) includes four additional sources in the 0.2-2 keV band, while the eROSITA All-Sky Catalog\footnote{https://erosita.mpe.mpg.de/dr1/AllSkySurveyData-dr1/} provides detections for two more sources and upper limits at the $3\sigma$ level for 13 additional sources in the 0.2-2.3 keV band (\citealt{Merloni24}; see \citealt{Tubin24} for details on calculation of upper limits). We exclude the hard X-ray bands from our analysis, as our study is limited to energies up to 2 keV in the observed frame. 

The X-ray flux densities at 1 keV have been calculated using WebPIMMS\footnote{https://heasarc.gsfc.nasa.gov/cgi-bin/Tools/w3pimms/w3pimms.pl}, assuming the quasar's emission in this band can be modeled as a power law with a photon index of 2 ($\Gamma = 2.0$), which is a typical value for luminous AGNs (e.g., \citealt{Reeves00}; \citealt{Piconcelli05}; \citealt{Shemmer05}). We use 1 keV as the normalization point, which is the standard reference energy (e.g., G07). In Table \ref{tab:table3}, the observed-frame 1 keV photometry is listed in units of $10^{-6}$ mJy for 23 GNIRS-DQS BAL quasars.

\section{Analysis \& Results} \label{sec:results}

\begin{figure*}
\figurenum{1}
\plotone{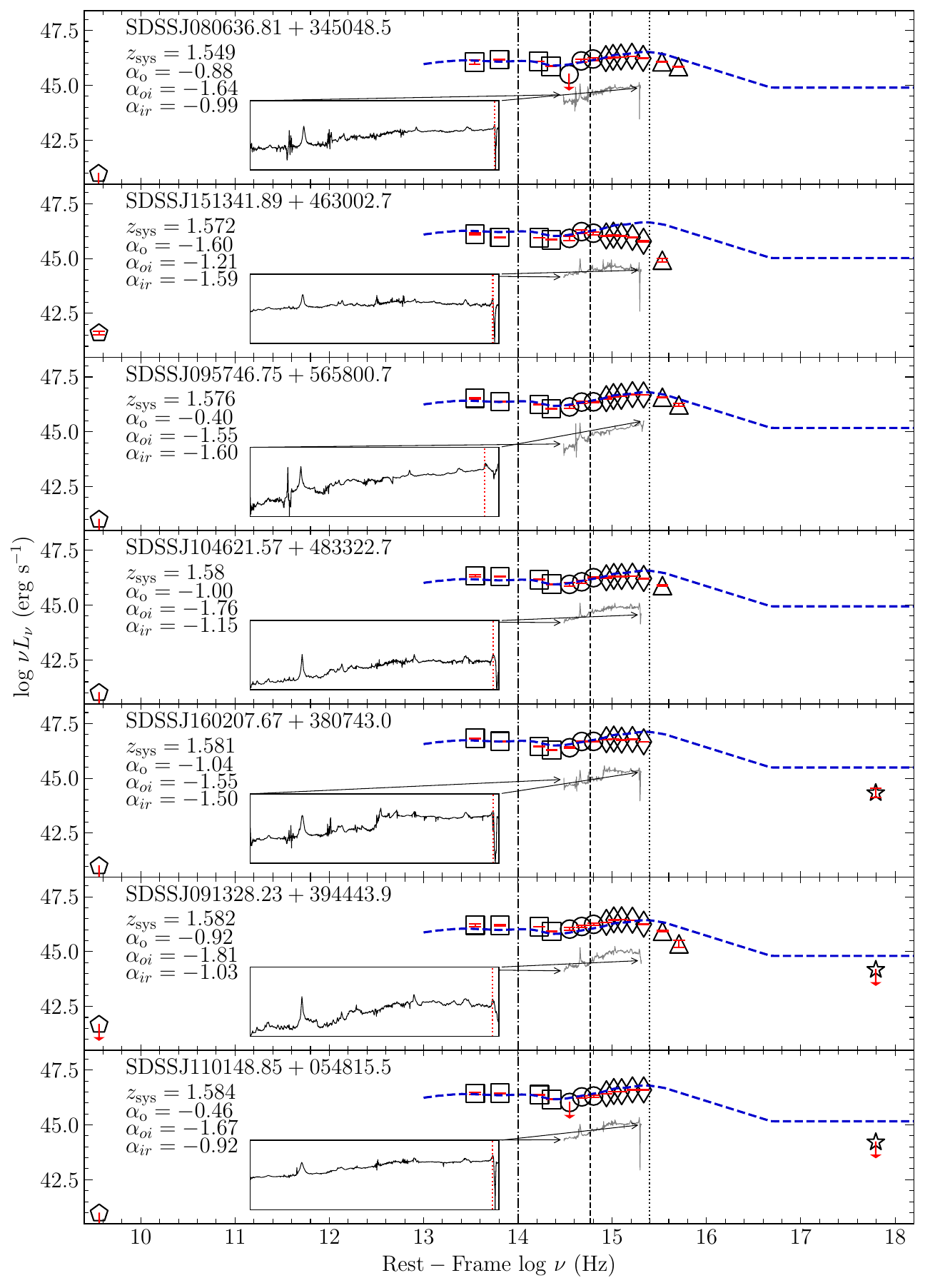}
\caption{SEDs for seven GNIRS-DQS BAL quasars listed in Table \ref{tab:table1} in order of increasing $z_{\rm sys}$. Symbols represent: Radio = pentagons, WISE = squares, 2MASS = circles, SDSS = diamonds, GALEX = triangles, and X-ray = stars. The high-luminosity composite SED (dashed blue line) from K13 has been scaled to the flux density at 5100 \AA\ (log($\nu$) = 14.77 Hz) in each source and over-plotted in each panel for reference. The gray spectrum in each panel, combined from M23 and \citet{Lyke20} data, is offset vertically for clarity; the inset panels zoom-in on these spectra with the \cfour~emission line marked by a red dotted line. Spectral regions at \hbox{$\lambda < 1200$ \AA} have been omitted due to the Ly$\alpha$ forest. The rest-frame wavelengths of 3 $\mu$m, 5100 \AA, and 1200 \AA\ are marked by dashed-dotted, dashed, and dotted lines, respectively. Objects are labeled with name, $z_{\rm sys}$, \alphao, \alphaoi, and \alphair~values. The remaining 58 SED plots are provided in Appendix \ref{sec:appendix}.}
\label{fig:Figure1}
\end{figure*}

The SEDs for the first seven GNIRS-DQS BAL quasars, sorted by increasing $z_{\rm sys}$, are presented in Figure \ref{fig:Figure1}. The remaining 58 quasar SED plots are provided in Appendix \ref{sec:appendix}. For reference, the composite high-luminosity SED from K13 (see Table 2 therein), spanning \hbox{log($\nu$) = 13.0–18.4 Hz} (30 $\mu$m to 10 keV), has been scaled and overplotted in each panel. Additionally, we have scaled and incorporated the spectra from M23 and \citet{Lyke20} for all 65 BAL quasars in each corresponding panel.

Many of the photometric points exhibit a drop blueward of \hbox{$\lambda_{\rm rest} \sim 1200$ \AA}, where intergalactic hydrogen causes significant attenuation of the quasar signal (i.e., the Ly$\alpha$ forest; \citealt{Lynds71}). We do not correct for host galaxy contamination because at high luminosities (i.e., log $(\lambda L_{\rm 5100\AA,total}) \geq 45 \rm~erg~s^{-1}$ ), the quasar completely outshines the host galaxy and no meaningful correction can be applied (e.g., \citealt{Shen11}).

\subsection{Mid-Infrared to X-ray BAL Quasar Composite SED} \label{subsec:comp}

\begin{figure*}
\figurenum{2}
\centering
\includegraphics[width=13.5cm]{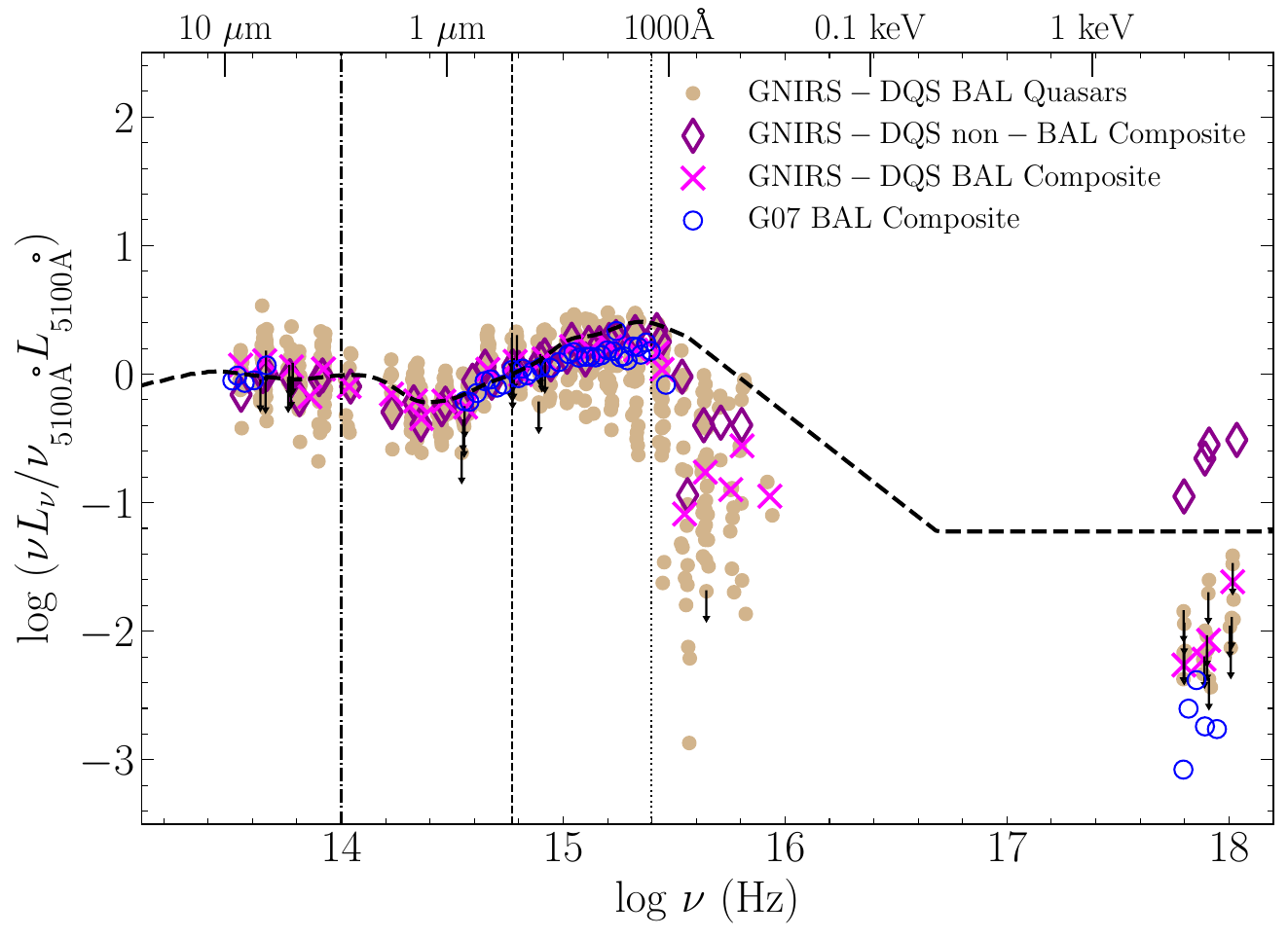}
\caption{GNIRS-DQS BAL quasars (brown filled circles) and their composite (magenta crosses), along with the GNIRS-DQS non-BAL composite (dark purple open diamonds) and LBQS BAL quasar composite (blue open circles). The method for constructing the BAL and non-BAL composite SED is described in Section \ref{subsec:comp}. Both the data and the K13 high-luminosity composite (black dashed curve) have been normalized to $\nu L_{\rm 5100\AA}$. The rest-frame wavelengths of \hbox{3 $\mu \rm m$ (log($\nu$) = 14.00 Hz)}, \hbox{5100 \AA\ (log($\nu$) = 14.77 Hz)}, and \hbox{1200 \AA\ (log($\nu$) = 15.40 Hz)} are indicated by dotted-dashed, dashed, and dotted lines, respectively. Upper limits are indicated by black arrows.} 
\label{fig:Figure2}
\end{figure*}

To comprehensively analyze the spectral properties of BAL and non-BAL quasars across a broad wavelength range, we construct composite SEDs using photometry from 24 $\mu$m to 1 keV, following the algorithm outlined by G07.

First, we normalize the SED of each quasar at a rest-frame wavelength of 5100~\AA\ (corresponding to a frequency of $\log \nu = 14.77$~Hz). We then apply a variable frequency window designed to include at least seven photometric data points for the objects in the sample, including upper limits, though some bins at the high-frequency end of the photometric coverage contain fewer points. Within each bin, we determine the median normalized luminosity to construct the composite SED.

While upper limits have minimal impact in the mid- and near-IR, and UV bands, their abundance in the X-ray regime for BAL quasars necessitates a more careful treatment to avoid overestimating the median luminosities. We apply Kaplan-Meier survival analysis  to the X-ray data for BAL quasars to account for the upper limits (e.g., \citealt{Kaplan58}; \citealt{Feigelson85}; \citealt{Schmitt85}). The median normalized luminosity is computed in each bin, thereby minimizing the impact of non-detections and reducing potential bias in the X-ray region (e.g., \citealt{Delhaize17}).

The normalized data for all BAL quasars with available photometry (including the 4 LoBAL quasars mentioned in Section \ref{sec:data}), and the non-BAL and BAL quasar composite are shown in Figure \ref{fig:Figure2}. We remove 16 RL quasars from the non-BAL composite and one RL BAL quasar from the BAL composite (see M23). From this point onward, all RL BAL and non-BAL quasars are excluded from our analysis unless explicitly stated otherwise. The LBQS BAL quasar composite from G07 and the high-luminosity SED composite from K13 normalized at $\nu L_{\rm 5100\AA}$ are also overplotted for reference. The GNIRS-DQS BAL quasar composite SED is given in Table \ref{tab:table4}.

To quantify the reddening in the GNIRS-DQS BAL quasar composite SED, we assume that all reddening occurs at each quasar's redshift and apply the SMC extinction curve (\citealt{Pei92}) to the composite SED. The SMC curve is chosen because the 2200 \AA\ bump present in the Large Magellanic Cloud and Milky Way extinction curves has rarely been detected in quasar dust (e.g., \citealt{Pitman00}). Using a $\chi^{2}$ minimization procedure over the wavelength range $1200~{\rm \AA} \leq \lambda \leq 10~\mu \rm m$, we redden the K13 composite SED until it matches the GNIRS-DQS BAL quasar composite SED. This process yields a color excess of $E(B-V) = 0.05~\pm~0.01$ for the GNIRS-DQS BAL quasar composite SED. Similarly to the BAL quasar composite, we find an $E(B-V) = 0.03~\pm~0.01$ for the GNIRS-DQS non-BAL composite SED.

We also examine the reddening of the GNIRS-DQS BAL quasar composite SED using the \citet{Gaskell04} extinction curve, which suggests flat UV extinction due to the relative lack of small dust grains in quasar environments. While this model produces an \hbox{$E(B-V) = 0.06~\pm~0.01$} for the GNIRS-DQS BAL quasar composite SED, consistent within the errors with the value obtained using the SMC curve, it does not provide an acceptable fit ($\chi^{2}$ of 7.5 compared to 1.1 for the SMC curve). Therefore, we adopt the SMC extinction curve to determine the best-fit $E(B-V)$ values for all 65 BAL (see Table \ref{tab:table1}) and 195 non-BAL quasars.

We compare the distributions of $E(B-V)$ values for 64 GNIRS-DQS BAL and 179 non-BAL quasars, shown in the top left panel of Figure \ref{fig:Figure3}. To test if $E(B-V)$ distributions for these samples differ significantly, we ran two-tailed Kolmogorov-Smirnov (K-S) and Anderson-Darling (A-D) tests on the parameter distributions shown in the top left panel of Figure \ref{fig:Figure3}. The A-D test exhibits greater sensitivity to differences in the tails of distributions. Significance thresholds for both tests were set at $p = 0.05$ and $p = 0.01$ to indicate rejection or failure to reject the null hypothesis, that BAL and non-BAL quasars originate from the same parent population, at the 95\% and 99\% confidence levels, respectively.

The results show that the null hypothesis—that BAL and non-BAL quasars come from the same parent population—is rejected at both the 95\% and 99\% confidence levels with both the K-S and A-D tests. While the means of the samples are consistent within their errors, the \hbox{K-S} and \hbox{A-D} tests indicate significant difference between GNIRS-DQS BAL and non-BAL quasars, with the former group showing more reddening. These results are consistent with previous studies and suggest that the increased reddening observed in the GNIRS-DQS BAL quasars may indicate a higher dust content or lines of sight with more dust in BAL quasars (e.g., \citealt{Sprayberry92}; \citealt{Reichard03}; \citealt{Trump06}; G07; see also \citealt{Ishibashi24}).

The average $E(B-V)$ value for GNIRS-DQS BAL quasars is higher than the average $E(B-V) = 0.02$ reported by \citet{Reichard03} for 180 HiBAL quasars, though their value for 34 LoBAL quasars is higher at 0.08, with a redshift range of $0.892 \leq z \leq 4.41$. Their method involved a two-parameter fit, adjusting both the power-law spectral index and $E(B-V)$ values in the 1355 \AA\ and 2250 \AA\ wavelength regions using a template composite spectrum. In contrast, \citet{Saccheo23} report an average $E(B-V)$ of 0.08 for 34 BAL quasars compared to 0.03 for 51 non-BAL quasars, similar to our procedure in finding $E(B-V)$ values for their sample of hyperluminous quasars ($L_{\rm bol} > 10^{47}~\rm erg~s^{-1}$) at $1.8 < z < 4.7$. The SMC extinction curve was also applied by \citet{Sprayberry92} for LoBAL quasars, who extrapolated the SMC extinction law from 1275 \AA\ down to 1000 \AA, and by \citet{Brotherton01}, for a RL LoBAL quasar with $z = 0.868$; both found $E(B-V) = 0.1$ for typical LoBAL quasars. Some differences between our $E(B-V)$ value and those reported in previous studies may stem from sample selection or differences in the wavelength ranges used for fitting. We also note that our GNIRS-DQS BAL quasar sample includes 4 LoBAL quasars, which may contribute to the higher average $E(B-V)$ value, as they are known to exhibit greater reddening than HiBAL quasars.

While the reddening observed in the GNIRS-DQS BAL and non-BAL composite SED is consistent within their errors, the distributions of individual \hbox{$E(B-V)$} values for BAL and non-BAL quasars reveal notable differences. Evidently, some BAL quasars exhibit greater reddening, which may not be fully captured by the composite SED. We also find no evidence of anomalous reddening in these sources, as their reddening behavior is consistent with SMC extinction. For a discussion on one example of anomalous reddening in a heavily reddened BAL quasar, Mrk 231, see \citet{Leighly14}. 

The GNIRS-DQS BAL sample also exhibits stronger \hbox{X-ray} absorption compared to the GNIRS-DQS non-BAL sample and the K13 composite, consistent with previous studies. While it is possible that some BAL quasars are intrinsically \hbox{X-ray} weak, the evidence largely supports the view that their \hbox{X-ray} weakness is primarily due to obscuration (e.g., \citealt{Gall02}, \citeyear{Gall06}; \citealt{Liu18}; \citealt{Wang22}; but see also \citealt{Teng14}; \citealt{Morabito14}). We also note the \hbox{X-ray} region for our BAL quasars is dominated by upper limits, whereas the LBQS BAL composite values include fewer upper limits and more robust detections (e.g., G07, see Table 3 therein).

\subsection{Optical Power-law Index} \label{subsec:alphao}

\begin{figure*}
\figurenum{3}
    \gridline{
        \fig{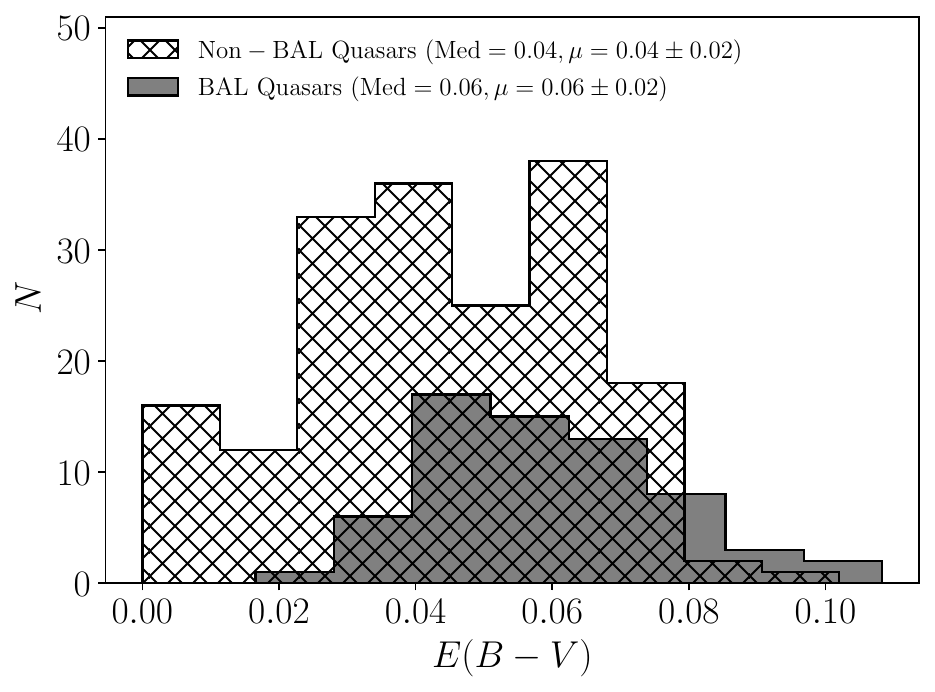}{0.45\textwidth}{}
        \fig{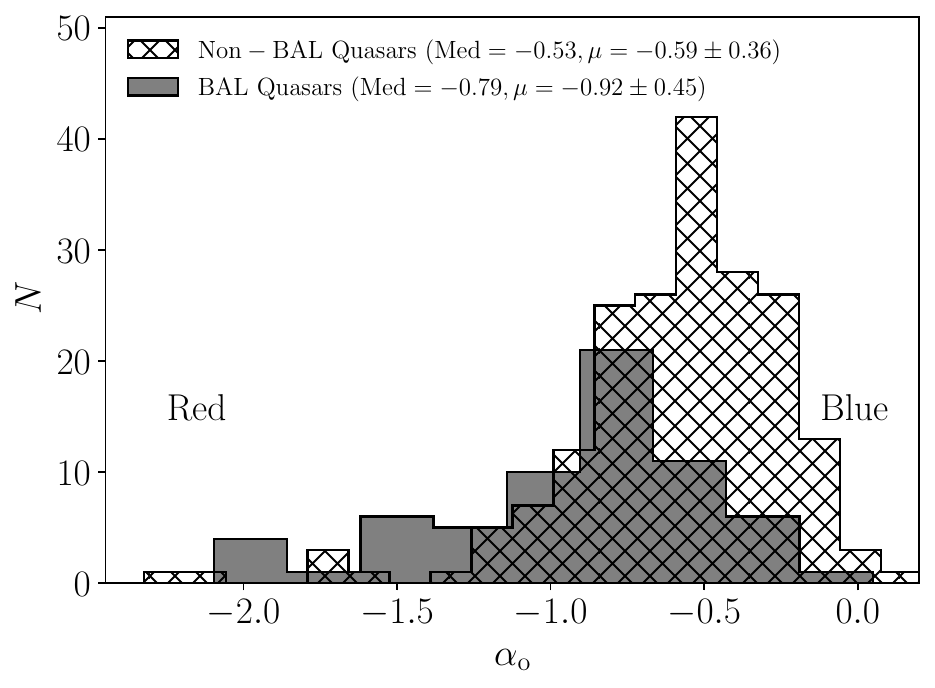}{0.45\textwidth}{}
    }
    \vspace{-2em}
    \gridline{
        \fig{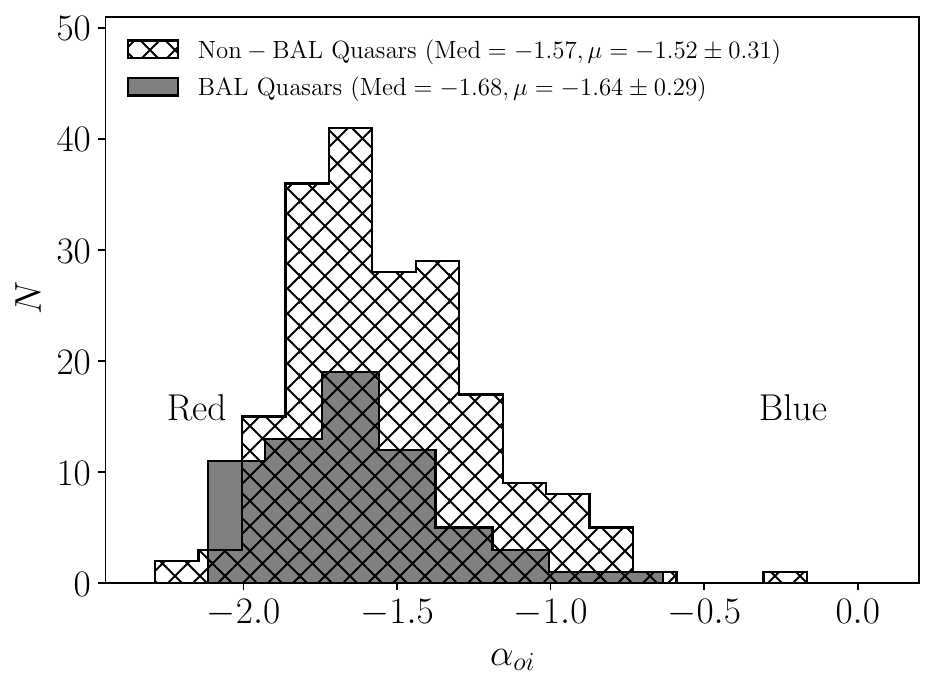}{0.45\textwidth}{}
        \fig{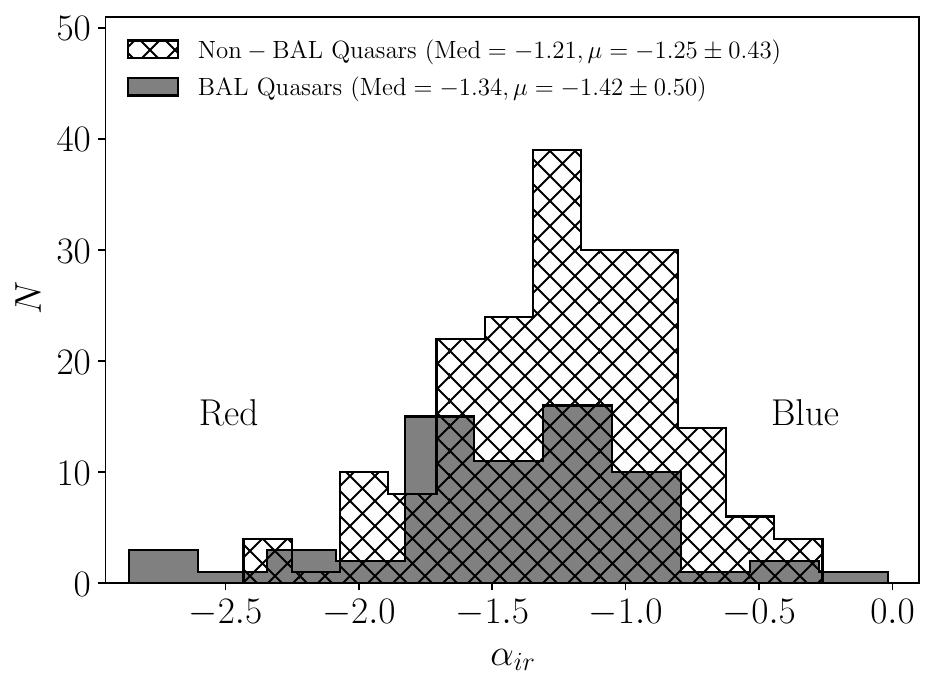}{0.45\textwidth}{}
    }\vspace{-2em}
    \caption{Distributions of $E(B-V)$ values (top left), \alphao~(top right), \alphaoi~(bottom left), and \alphair~(bottom right) for BAL (gray-filled), and non-BAL (hatched) GNIRS-DQS quasar samples. We apply adaptive binning using the Freedman-Diaconis rule for all distributions to ensure optimal bin sizes. Higher $E(B-V)$ values correspond to greater reddening, while more negative values of \alphao, \alphaoi, and \alphair~indicate redder continua in the UV-optical, near-IR, and mid-IR regions, respectively. Median (Med), and mean ($\mu$) values with the standard deviation are indicated in the legends. GNIRS-DQS BAL quasars show a significantly redder distribution in the UV-optical, yet not in the near- and mid-IR continuum compared to non-BAL quasars.}
    \label{fig:Figure3}
\end{figure*}

To characterize the color of the optical continuum, we fit a power-law model ($L_{\nu}\propto \nu^{\alpha_{\rm o}}$), where \alphao~is the optical power-law index, to the photometric data between rest-frame 1200 and 5000 \AA\ using 2MASS and SDSS data. This spectral region is typically dominated by the quasar emission, allowing us to measure the shape of the optical continuum arising from the accretion disk. The \alphao~parameter can vary intrinsically from one quasar to another and may become more negative due to dust reddening. Wavelengths \hbox{$\lambda < 1200$ \AA} are to be avoided for the redshift range of our sample, as the intervening Ly$\alpha$ forest reduces the flux significantly.

We compare the distribution of \alphao~for 64 BAL quasars, and 179 non-BAL quasars presented in the top right panel of Figure \ref{fig:Figure3}. Similar to Section \ref{subsec:comp}, we ran two-tailed K-S and \hbox{A-D} tests to evaluate the null hypothesis, that BAL and non-BAL quasars come from the same parent population, at both the 95\% and 99\% significance levels. The results show that the null hypothesis is rejected at both the 95\% and 99\% confidence levels with both the K-S and A-D tests. This result is consistent with result of the $E(B-V)$ distributions of BAL and non-BAL quasars.

Our analysis reveals that GNIRS-DQS BAL quasars have, as a group, a more negative optical power-law index than non-BAL quasars, indicating redder UV-optical spectra, which is consistent with previous studies and likely results from higher dust extinction (e.g., \citealt{Reichard03}; \citealt{Trump06}; G07; \citealt{Dai08}; \citealt{K15}; \citealt{Ishibashi24}). At the blue (more positive) end, the distribution is roughly Gaussian, with a long tail extending towards the red (more negative) end. Specifically, the median values of \alphao\ are -0.53 for non-BAL quasars and -0.79 for BAL quasars, reflecting this spectral distinction.

We perform Spearman rank correlations between \alphao~for GNIRS-DQS BAL and non-BAL quasars and various parameters, including rest-frame equivalent width (EW) of the \othree~$\lambda 5007$ emission line, EW of \hb~$\lambda 4861$ emission line, full-width at half maximum (FWHM) intensity of \hb, black hole mass ($M_{\rm BH}$), Eddington ratio ($L/L_{\rm Edd}$), and $R_{\rm Fe{II}}$ (ratio of EW(Fe~{\sc ii})/EW(H$\beta$)), using data from M23. The BAL quasar analysis also includes the BI and AI values from \citet{Ahmed24}. We find no significant correlations between \alphao~and any of these parameters.

\subsection{Optical-Near-Infrared Continuum Shape} \label{subsec:alphaoi}

In addition to the optical power-law index, we also compute the point-to-point flux density slope, \alphaoi, between the flux densities at 5100 \AA\ and 3 $\mu$m defined in \citet{Leighly24} as

\begin{equation}
    {a_{oi}} = \frac{{\log (F_{\nu, 3~\mu m})} - {\log (F_{\nu,\rm 5100~\AA})}}{{\log (\nu_{3~\mu m})} - {\log (\nu_{\rm 5100~\AA})}}.
\end{equation}

\hfill

\noindent The value of \alphaoi~can depend on three quasar properties: reddening, the intrinsic shape of the quasar SED, and the strength of the torus emission. It may be less negative (bluer) due to weaker torus emission. Conversely, it may be more negative (redder) because the torus emission is strong, or because there is significant reddening.

The flux density at 5100 \AA\ for BAL and non-BAL quasars is taken from M23. The flux density at rest-frame \hbox{3 $\mu$m} is estimated from scaling the quasar composite SED of K13 to the WISE photometry data and interpolating the flux density to \hbox{3 $\mu$m}.

The distributions of \alphaoi~for GNIRS-DQS BAL and non-BAL quasars are presented in the bottom left panel of Figure \ref{fig:Figure3}. We ran two-tailed K-S and \hbox{A-D} tests to evaluate the null hypothesis, that BAL and non-BAL quasars come from the same parent population, at both the 95\% and 99\% significance levels. The K-S test result reveals that the null hypothesis is not rejected at either significance level. In contrast, the A-D test results reject the null hypothesis at the 95\% significance level but not at the 99\% level. 

These findings suggest that BAL quasars exhibit similar torus emission, indicative of hot dust signatures, compared to the non-BAL quasar sample (see also \citealt{Leighly24}). We find that the near-IR SED shapes of GNIRS-DQS BAL and non-BAL quasars do not differ significantly. Similar to Section \ref{subsec:alphao}, Spearman rank correlations between \alphaoi~and other parameters, such as EW(\othree), EW(\hb), FWHM(\hb), $M_{\rm BH}$, $L/L_{\rm Edd}$, $R_{\rm Fe{II}}$, BI and AI, reveal no significant correlations for either BAL or non-BAL quasars.

Overall, we find no evidence of significant excess in near-IR emission for GNIRS-DQS BAL quasars compared to their non-BAL counterparts. Our result is inconsistent with the results of \citet{Dipompeo13}, who find a significant excess in the mid- to near-IR luminosities of BAL quasars. However, their sample predominantly includes RL BAL quasars, which are known to exhibit increased reddening \citep{Brotherton01b}. Their result was based on a comparison of the luminosity ratio at rest-frame wavelengths of \hbox{4 $\mu \rm m$} and \hbox{2500 \AA}, whereas our comparison is made using Equation 1 at rest-frame wavelengths of \hbox{3 $\mu \rm m$} and \hbox{5100 \AA}. To directly compare with their sample, we interpolate all luminosities to \hbox{4 $\mu$m}, in a similar manner to \hbox{3 $\mu$m}, and compare the luminosity ratio at rest-frame wavelengths of \hbox{4 $\mu \rm m$} and \hbox{2500 \AA}. Again, the null hypothesis is rejected at only the 95\% significance level using the A-D test, but not the K-S test, and our findings do not indicate a significant difference between the two populations of quasars.

When interpreting the \alphaoi~values, we consider the impact of variability, particularly since the optical continuum at 5100 \AA\ and the IR emission at 3 $\mu$m may not have been observed simultaneously. Quasars are known to vary in their emission, with a time lag between changes in the optical continuum and corresponding IR emission due to dust reprocessing (e.g., \citealt{Kishimoto07}). However, this lag is unlikely to introduce significant asymmetry in the optical-IR emission distribution, and there is no strong evidence that BAL quasars exhibit greater variability in their continuum emission compared to non-BAL quasars (e.g., \citealt{Gibson08}).

\subsection{Mid-Infrared Power-law Index} \label{subsec:alphair}

Since the mid-IR (1-10 $\mu$m) continuum spectrum is primarily dominated by thermal dust emission, we perform power-law fits ($L_{\nu} \propto \nu^{\alpha_{ir}}$), where \alphair~is the mid-IR power-law index, to the photometric data in the rest-frame 1.8-10 $\mu$m region (e.g., \citealt{Gall07b}). This spectral range was chosen to maximize photometric coverage that is primarily probing the warm dust emission. We note that silicate emission around $\sim 10~\mu \rm m$ is typically weak in most AGN spectra and is not expected to significantly affect results in this spectral region (e.g., \citealt{Hill14}, see Figure 3 therein).

The \alphair~distributions for 179 non-BAL, and 64 BAL GNIRS-DQS quasars are shown in bottom right panel of Figure \ref{fig:Figure3}. We performed K-S and A-D tests on these distributions to assess whether BAL and non-BAL quasars originate from the same parent population, using 95\% and 99\% confidence levels. The K-S test indicates that the null hypothesis cannot be rejected at either significance level, while the A-D test results show rejection of the null hypothesis at the 95\%, but not at the 99\%, confidence level.

Similar to Sections \ref{subsec:alphao} and \ref{subsec:alphaoi}, no significant Spearman-rank correlations were found between \alphair~and other basic parameters (EW(\othree), EW(\hb), FWHM(\hb), $M_{\rm BH}$, $L/L_{\rm Edd}$, $R_{\rm Fe{II}}$, BI and AI) for BAL and non-BAL quasars. Ultimately, we find no compelling evidence of significant differences in the broad-band mid-IR emission between BAL and non-BAL quasars.

The fact that BAL quasars are reddened in the UV-optical and exhibit overall no significant difference in the near- and mid-IR spectra suggests that the warm dust emission is similar to non-BAL quasars. This result is in agreement with the findings of G07, who compare the mid-IR luminosity at $8~\mu \rm m$ to the optical luminosity at $5000~\rm \AA$. They find no significant differences in the mid-IR relative to optical power between BAL and non-BAL quasars of similar luminosity, further supporting the idea that the underlying torus properties and dust reprocessing mechanisms are not substantially different in these two populations.

\subsection{Bolometric Luminosity \& Corrections} \label{subsec:BC}

Since quasar photometry is usually restricted to a few bandpasses, estimating key parameters such as black hole masses and Eddington ratios requires an accurate determination of the integrated bolometric luminosity. Thus, we determine the bolometric luminosity of the BAL quasars by integrating the area under the SED as

\begin{equation}
    {L_{\rm bol}} = \int_{0}^{\infty} L_{\nu}dv = \int_{-\infty}^{\infty} {\rm ln(10)}~\nu L_{\nu}~d~{\rm log}(v).
\end{equation}

\hfill

As discussed by \citet{Marconi04}, the full observed SED includes reprocessed light, therefore, the resulting SED does not accurately reflect the intrinsic emission. It should be noted that a significant portion of the IR radiation ($\sim1~\mu$m – $30~\mu$m) is thought to originate from the torus (e.g., \citealt{Krolik88}; \citealt{Elitzur06}), where high-energy photons are absorbed and re-radiated. This reprocessing can lead to double-counting in $L_{\rm bol}$, as both IR and hard X-ray (energies larger than 2 keV) emissions include contributions from reprocessed disk radiation. To avoid counting the same contribution twice, several authors (e.g., \citealt{Marconi04}; \citealt{Nemmen10}; \citealt{Runnoe12a}) limit the integral over the range of $1~\mu \rm m~to~8~\rm keV$. Additionally, dust reddening in the UV-optical can result in an underestimation of the observed luminosity relative to the true intrinsic value.

Therefore, we follow the methodology of K13 to estimate the bolometric luminosity from $1~\mu$m to 2 keV. While K13 follows an agnostic approach and gives $L_{\rm bol}$ values of multiple integrated regions (see Table 3 therein), they recommend the range of $1~\mu$m to 2 keV (which avoids the issue of IR and hard X-ray double counting). Following their approach, we do not correct for non-isotropic emission in our calculations; however, taking anisotropy into account, \citet{Runnoe12a} suggest scaling the bolometric luminosities by 0.75 when calculating the bolometric luminosity over the range of $1~\mu$m to 8 keV.

In regions of limited photometric coverage, the K13 composite SED was used to complete the wavelength coverage. When X-ray detections are available, we incorporate the data points at 2 keV directly into the integration. In the absence of X-ray data, we extrapolate from the K13 composite SED at 2 keV. The bolometric corrections (BC) were determined at 5100 \AA, and 3 $\mu$m using

\begin{equation}
    {\rm BC_{\nu}} = \frac{L_{\rm bol}} {\nu L_{\nu}}.
\end{equation}

\citet{Runnoe12a} found that nonlinear relationships in the form of \hbox{log($L_{\rm bol}$) = $A~{\rm log}(L_{\lambda})+B$} provided a better representation of the data (see also \citealt{Nemmen10}). In the case of 5100 \AA, \citet{Runnoe12a} found \hbox{log($L_{\rm bol}$) $= (0.91 \pm 0.04)~{\rm log}(L_{5100 \rm \AA})+ (4.89 \pm 1.66)$}, whereas K13 give a constant BC of $4.33 \pm 1.29$. In the case of 3 $\mu$m, \citet{Runnoe12b} report BC of \hbox{log($L_{\rm bol}$) $= (0.92 \pm 0.08)~{\rm log}(L_{3 \rm \mu m}) + (4.54 \pm 3.42)$}, while K13 do not provide a BC at 3 $\mu$m.

In addition, \citet{Elvis94} and \citet{Richards06} report BCs at 5100 \AA\ of \hbox{11.8 $\pm$ 12.9} and \hbox{10.3 $\pm$ 2.1}, respectively. However, it should be noted that they compute $L_{\rm bol}$ over the range of $\sim30~\mu$m to $10$ keV. \citet{Gall07b} provides BC at \hbox{3 $\mu$m} of \hbox{8.59 $\pm$ 3.30} from mid-IR SEDs of 234 radio-quiet quasars originally presented by \citet{Richards06}. 

The $L_{\rm bol}$ values for all 65 GNIRS-DQS BAL quasars are provided in Table \ref{tab:table1}. We find an average BC of \hbox{2.40 $\pm$ 0.65} at 5100 \AA\ and \hbox{2.76 $\pm$ 1.54} at 3 $\mu$m. For comparison, we compute BCs for GNIRS-DQS non-BAL quasars as well, following the same methodology, and find an average BC of \hbox{4.56 $\pm$ 1.32} at 5100 \AA\ and \hbox{6.21 $\pm$ 3.58} at 3 $\mu$m. The BCs for the non-BAL quasars in our sample are consistent with previous studies. However, the average BC values at both 5100 \AA\ and 3 $\mu$m are lower in the BAL quasar sample\footnote{We note that these are not intrinsic luminosities, as we do not correct for reddening or the X-ray absorption.}, likely due to their greater reddening in the UV-optical region (see Section \ref{subsec:alphao}) and weak X-ray emission (see Figure \ref{fig:Figure2}).

\subsection{Covering Fractions} \label{subsec:CF}

While our results show no significant differences in the continuum properties of BAL and non-BAL quasars, aside from the UV-optical regime, we further extend our analysis by estimating the torus covering fractions for both populations to assess whether they exhibit any significant differences.

In previous studies, various proxies have been used to estimate the covering fraction of the obscuring torus in AGNs. Some approaches employ the ratio between monochromatic luminosities at specific IR and UV-optical wavelengths (e.g., \citealt{Maiolino07}; \citealt{Treister08}; \citealt{Trefoloni24}), while others integrate over broader wavelength ranges to capture the full spectral output (e.g., \citealt{Cao05}; \citealt{Gu13}). 

We use observational proxies based on the combination of UV-optical and near-IR luminosities. Specifically, following the approach in \citet{Maiolino07}, we estimate the torus covering fraction using

\begin{equation}
    {\rm Covering~Fraction} = (f) ~\frac{\nu L_{\rm 3\mu m}} {\nu L_{5100 \rm \AA}},
\end{equation}

\noindent where $f$ represents the ratio of the average BC values at \hbox{3 $\mu$m} and 5100 \AA\ for both BAL and non-BAL quasars, as discussed in Section \ref{subsec:BC}. 

The average covering fractions for BAL and non-BAL quasars are $1.20~\pm~0.54$, and $1.19~\pm~0.51$, respectively. We conduct a two-sample $t$-test to assess the null hypothesis that BAL and non-BAL quasars have the same average covering fractions at the 95\% and 99\% confidence levels. The $t$-test result indicates that the null hypothesis is not rejected at either the 95\% or 99\% confidence levels. This result suggests that the mean covering fractions of the obscuring torus for BAL and non-BAL quasars are not significantly different, consistent with our findings in Section \ref{subsec:alphaoi}. While these results contrast with \citet{Dipompeo13}, who found higher covering fractions for BAL quasars, the difference they report is only at the 10\% level.

Our results indicate that GNIRS-DQS BAL and non-BAL quasars have average torus covering fractions that are consistent within the errors. The lack of a significant difference in covering fractions is notable in the context of the disk-wind model, which suggests that BAL quasars are observed along sightlines intersecting winds (e.g., \citealt{Hamann93}). \citet{Ahmed24} demonstrate that a threshold luminosity of \hbox{$\lambda L_{2500} \gtrsim 10^{45}$ erg s$^{-1}$} is necessary for outflow production. Even in the most luminous quasars, though, the BAL fraction remains below $\sim$40\%, likely due to orientation constraints.

If BAL quasars were to have higher covering fractions, this would also suggest more efficient reprocessing of UV-optical radiation into IR emission, implying that BAL quasars should appear brighter in the near- and mid-IR (e.g., \citealt{Dipompeo13}). However, our findings do not support this expectation (see Sections \ref{subsec:alphaoi} and \ref{subsec:alphair}). This discrepancy highlights a key challenge in the disk-wind paradigm: reconciling covering fraction differences, viewing angle effects, and the lack of a detectable IR excess. It remains unclear whether the inclination angles of BAL and non-BAL quasars differ enough to explain these results, or if additional factors, such as wind geometry or dust properties, contribute to the observed behavior.

\section{Summary and Conclusions} \label{sec:conclusion}

We present a comprehensive SED analysis of 65 high-luminosity, high-redshift GNIRS-DQS BAL quasars. The GNIRS-DQS dataset provides the largest uniform inventory of rest-frame optical spectral properties of luminous quasars at high redshift. This dataset was used to construct a new BAL quasar composite SED with coverage from rest-frame 24 $\mu$m to 1 keV. We also present bolometric corrections at \hbox{5100 \AA} and \hbox{3 $\mu$m}, derived from the integrated light across the \hbox{1 $\mu$m} to \hbox{2 keV} range. 

Our findings reveal that BAL quasars in the GNIRS-DQS sample exhibit significantly redder UV-optical continua compared to their non-BAL counterparts, a difference likely driven by increased dust extinction (e.g., \citealt{Reichard03}; \citealt{Trump06}; \citealt{Gibson09}; \citealt{Urrutia09}; \citealt{Allen11}). Additionally, BAL quasars show no significant differences in their near- and mid-IR continuum properties compared to non-BAL quasars, a result consistent with previous studies (e.g., G07; \citealt{Leighly24}; but see also \citealt{Dipompeo13}). As previously established, the most notable difference between BAL and non-BAL quasar SEDs is in the \hbox{X-ray} regime, where BAL quasars are weak X-ray emitters, evident with GNIRS-DQS BAL sources as well. While some BAL quasars may be intrinsically X-ray weak, evidence generally suggests that X-ray weakness in these sources is primarily due to obscuration (e.g., \citealt{Gall02}, \citeyear{Gall06}; \citealt{Liu18}; \citealt{Wang22}; but, see also \citealt{Teng14}; \citealt{Morabito14}). 

These observations may be interpreted within the framework of the disk-wind paradigm, where most, if not all, luminous quasars possess a BAL region. However, the detectability of BAL features depends on orientation. According to \citet{Ahmed24}, they assume that all high-luminosity quasars, such as those in GNIRS-DQS, possess winds driven by their high accretion rates. Yet, the observed fraction of BAL quasars remains limited to approximately 25\% (see Figure 5 therein), as the winds are detectable only along specific lines of sight where the outflow intersects the observer's view, resulting in increased obscuration and reddening.

These outflows, driven by radiation pressure on UV resonance lines, can carry dust, forming dusty winds that contribute to the IR emission (e.g., \citealt{Murray95}; \citealt{Elvis2000}; \citealt{Gall15}; see also \citealt{Ishibashi24}). The dust within these outflows may also cause reddening, which likely accounts for the diminished X-ray and UV-optical emission frequently observed in BAL quasars. Comparable torus covering fractions for BAL and non-BAL quasars (see Section \ref{subsec:CF}) suggest that the structure and obscuration properties of the torus are similar in both populations, indicating that differences in their appearance are unlikely to arise from intrinsic differences in the torus itself. Despite expectations that BAL quasars should exhibit stronger IR emission (e.g., \citealt{DiPompeo12}), our data show no significant IR excess. The IR strength likely depends as much on orientation as on the amount of dust present (e.g., \citealt{Proga00}), as viewing angles influence how much dust near the torus is visible. In addition, the inclination angle of outflows depends on the wind's launching radius, further complicating the relationship between orientation and IR brightness (e.g., \citealt{Everett05}).

Overall, our results show no compelling evidence for significant differences in the SEDs of BAL and non-BAL quasars with comparable luminosities, aside from the reddening of the UV-optical continuum in the former group. These findings align with recent studies that both groups also exhibit similar rest-frame optical emission-line properties (e.g., \citealt{Ahmed24}; \citealt{Temple24}). The consistency in these results suggests that the underlying physical processes shaping their emission are similar, and any observed differences are likely attributable to orientation or other external factors rather than intrinsic properties.

While current research shows that BAL and non-BAL quasars have similar luminosities, black hole masses, Eddington ratios, and near-to-mid-IR continuum properties, the role of orientation remains unresolved. Future studies would benefit from employing proxies for quasar orientation, such as radio morphology, polarization, or emission-line anisotropies, to disentangle the impact of viewing angles on observed properties (e.g., \citealt{DiPompeo12}; \citealt{Dipompeo13}; \citealt{Richards21}; \citealt{Nair22}).  Detailed modeling of disk-wind geometries and their dependence on luminosity, wind strength, and launching radius is needed to further refine our understanding of the relationship between BAL and non-BAL quasars. Additionally, identifying reliable proxies that directly connect reddening to orientation could be significant, offering deeper insights into the complex interplay between outflows, obscuration, and orientation in quasar systems.

\raggedbottom
\begin{acknowledgments}

This work is supported by the Natural Science and Engineering Research Council (NSERC) RGPIN-2021-04157 and the Western Research Leadership Chair. We thank an anonymous referee for constructive comments that improved this manuscript. This research has made use of the NASA/IPAC Extra-galactic Database (NED), which is operated by the Jet Propulsion Laboratory, California Institute of Technology, under contract with the National Aeronautics and Space Administration. This research has also made use of data obtained from the 4XMM XMM-Newton Serendipitous Source Catalog compiled by the 10 institutes of the XMM-Newton Survey Science Centre selected by ESA. H.A. acknowledges support by the University of North Texas Graduate Research Experiences Abroad Travel Grant.
\end{acknowledgments}

\software{matplotlib \citep{Hunter07},  
          numpy \citep{van11}; \citep{Harris20}, 
          pandas \citep{Mckinney10},
          scipy \citep{Virtanen20},
          scikit-learn \citep{Pedregosa11}
          }
\pagebreak

\bibliography{references}{}
\bibliographystyle{aasjournalv7}

\begin{longrotatetable}
\begin{deluxetable}{lccccccccccccccc}
\tablenum{2}
\tablecaption{Observed-Frame Radio and Infrared Photometry of GNIRS-DQS BAL Quasars \label{tab:table2}}
\tablecolumns{9}
\setlength{\tabcolsep}{6 pt}
\tablewidth{0.5 cm}
\tabletypesize{\scriptsize}
\tablehead{
\colhead{Quasar} & \colhead{Radio$^{a}$} & \multicolumn{4}{c}{WISE$^{b}$} & \multicolumn{3}{c}{2MASS$^{c}$} \\
\cmidrule(lr){3-6}
\cmidrule(lr){7-9}
 \colhead{(SDSS J)} & \colhead{1.4 GHz} & \colhead{\textit{W1} - $3.4~\mu$m}  & \colhead{\textit{W2}- $4.6~\mu$m} & \colhead{\textit{W3} - $12.0~\mu$m}& \colhead{\textit{W4} - $22.0~\mu$m} & \colhead{\textit{J} - $1.25~\mu$m} & \colhead{\textit{H} - $1.65~\mu$m} & \colhead{\textit{K} - $2.16~\mu$m}\
}

\startdata
$001249.89+285552.6$ & \nodata & 0.361 $\pm$ 0.011 & 0.417 $\pm$ 0.016 & 2.253 $\pm$ 0.156 & $<4.180$ & 0.397 $\pm$ 0.045 & 0.528 $\pm$ 0.068 & 0.407 $\pm$ 0.080\\
$001355.10-012304.0$ & $<0.432$ & 0.367 $\pm$ 0.011 & 0.416 $\pm$ 0.016 & 2.199 $\pm$ 0.168 & 6.516 $\pm$ 1.056 & 0.328 $\pm$ 0.043 & 0.389 $\pm$ 0.053 & 0.419 $\pm$ 0.062\\
$004613.54+010425.7$ & 3.300 $\pm$ 0.400$^{*}$ & 0.424 $\pm$ 0.013 & 0.475 $\pm$ 0.018 & 1.420 $\pm$ 0.201 & $<5.649$ & 0.421 $\pm$ 0.047 & 0.468 $\pm$ 0.068 & 0.629 $\pm$ 0.084\\
$013012.36+153157.9$ & 20.070 $\pm$ 0.700$^{*}$ & 0.557 $\pm$ 0.015 & 0.628 $\pm$ 0.021 & 1.919 $\pm$ 0.125 & 2.935 $\pm$ 0.808 & 0.424 $\pm$ 0.044 & 0.480 $\pm$ 0.054 & 0.834 $\pm$ 0.071\\
$013652.52+122501.5$ & $1.740 \pm 0.699$ & 0.589 $\pm$ 0.016 & 1.231 $\pm$ 0.031 & 5.424 $\pm$ 0.180 & 10.744 $\pm$ 0.950 & 0.351 $\pm$ 0.046 & 0.497 $\pm$ 0.065 & 0.819 $\pm$ 0.078\\
\enddata
\tablecomments{All flux densities are in units of mJy. (This table is available in its entirety in machine-readable format.)}
\tablenotetext{a}{Radio photometry compiled from the FIRST (\citealt{White97}) and NVSS (\citealp{Condon98}) catalogs (Sources marked with * have coverage from NVSS).}
\tablenotetext{b}{Mid-IR photometry from the WISE final data release (\citealt{Wright10}).}
\tablenotetext{c}{Near-IR photometry from the 2MASS catalog (\citealt{Skrutskie06}).}
\end{deluxetable}
\end{longrotatetable}

\begin{longrotatetable}
\begin{deluxetable*}{lcccccccccccc}
\tablenum{3}
\tablecaption{Observed-Frame Optical, Ultraviolet, and X-ray Photometry of GNIRS-DQS BAL Quasars \label{tab:table3}}
\tablecolumns{9}
\setlength{\tabcolsep}{6 pt}
\tablewidth{0.5 cm}
\tabletypesize{\scriptsize}
\tablehead{
\colhead{Quasar} & \multicolumn{5}{c}{SDSS$^{a}$} & \multicolumn{2}{c}{GALEX$^{b}$} & \colhead{X-ray$^{c}$}\\
\cmidrule(lr){2-6}
\cmidrule(lr){7-8}
\cmidrule(lr){9-10}
 \colhead{(SDSS J)} & \colhead{\textit{u} - $\rm 3587~\AA$} & \colhead{\textit{g} - $\rm 4717~\AA$}& \colhead{\textit{r} - $\rm 6165~\AA$} & \colhead{\textit{i} - $\rm 7476~\AA$} & \colhead{\textit{z} - $\rm 8923~\AA$} & \colhead{\textit{FUV} - $\rm 1516~\AA$} & \colhead{\textit{NUV} - $\rm 2267~\AA$} & \colhead{1 keV}\
}

\startdata
$001249.89+285552.6$ & 0.002 $\pm$ 0.001 & 0.090 $\pm$ 0.002 & 0.161 $\pm$ 0.003 & 0.207 $\pm$ 0.003 & 0.234 $\pm$ 0.007 & \nodata & \nodata & \nodata \\
$001355.10-012304.0$ & 0.0003 $\pm$ 0.0002 & 0.114 $\pm$ 0.002 & 0.195 $\pm$ 0.003 & 0.239 $\pm$ 0.004 & 0.265 $\pm$ 0.005 & \nodata & 0.002 $\pm$ 0.001 & \nodata \\
$004613.54+010425.7$ & 0.141 $\pm$ 0.003 & 0.181 $\pm$ 0.003 & 0.228 $\pm$ 0.003 & 0.267 $\pm$ 0.004 & 0.311 $\pm$ 0.007 & 0.011 $\pm$ 0.001 & 0.029 $\pm$ 0.001 & 2.166 $\pm$ 0.793 \\
$013012.36+153157.9$ & 0.040 $\pm$ 0.002 & 0.129 $\pm$ 0.006 & 0.252 $\pm$ 0.005 & 0.342 $\pm$ 0.006 & 0.455 $\pm$ 0.008 & \nodata & \nodata & \nodata \\
$013652.52+122501.5$ & 0.084 $\pm$ 0.003 & 0.179 $\pm$ 0.004 & 0.230 $\pm$ 0.003 & 0.263 $\pm$ 0.004 & 0.347 $\pm$ 0.007 & \nodata & 0.004 $\pm$ 0.001 & \nodata \\
\enddata
\tablecomments{All flux densities are in units of mJy, except for X-ray fluxes which are in units of $10^{-6}$ mJy. (This table is available in its entirety in machine-readable format.)}
\tablenotetext{a}{Optical data are dereddened PSF fluxes from SDSS DR16 \citep[Table D1, Column 92]{Lyke20}.}
\tablenotetext{b}{UV photometry corrected for Galactic extinction from GALEX GR6 (\citealt{Martin05}).}
\tablenotetext{c}{X-ray data taken from the Chandra Point Source Catalog (\citealt{Evans10}), 4XMM XMM-Newton Serendipitous Source Catalog (\citealt{Webb20}), and eROSITA All-Sky Catalog (\citealt{Merloni24}).}
\end{deluxetable*}
\end{longrotatetable}

\clearpage

\startlongtable
\begin{deluxetable}{lcccccccccccccccccc}
\tablenum{4}
\tablecaption{GNIRS-DQS BAL Quasar Composite SED \label{tab:table4}}
\tablecolumns{3}
\setlength{\tabcolsep}{2.8 pt}
\tablewidth{0.5 cm}
\tabletypesize{\scriptsize}
\tablehead{
 \colhead{$\nu^{a}$} & \colhead{${L_{\nu}/L_{\rm 5100\AA}}^{b}$}  & \colhead{Number$^{c}$} \
}
\startdata
13.547  &  0.069   &  12 \\
13.655  &  0.103   &  41 \\
13.776  &  0.057   &  19 \\
13.859  & -0.176   &  10 \\
13.920  &  0.040   &  36 \\
14.035  & -0.097   &  12 \\
14.227  & -0.155   &  12 \\
14.334  & -0.209   &  41 \\
14.362  & -0.345   &  15 \\
14.462  & -0.210   &  50 \\
14.561  & -0.256   &  24 \\
14.661  &  0.037   &  53 \\
14.784  &  0.098   &  56 \\
14.808  & -0.004   &  15 \\
14.914  &  0.119   &  59 \\
15.037  &  0.215   &  65 \\
15.124  &  0.186   &  59 \\
15.212  &  0.214   &  71 \\
15.326  &  0.186   &  64 \\
15.426  &  0.166   &  14 \\
15.446  &  0.037   &  40 \\
15.547  & -1.089   &  20 \\
15.640  & -0.763   &  33 \\
15.755  & -0.899   &  12 \\
15.931  & -0.558   &   9 \\
15.806  & -0.950   &   7 \\  
17.797 & -2.263 & 6\\
17.887 & -2.218 & 4\\
17.908 & -2.071 & 6\\
18.017 & -1.615 & 7\\
\enddata
\tablenotetext{a}{Logarithm of the rest-frame frequency (Hz) in the center of the bin.}
\tablenotetext{b}{Logarithm of the ratio of the monochromatic luminosity at the given frequency to that at 5100 \AA.}
\tablenotetext{c}{Number of photometric data points contributing to the respective frequency bin.}
\end{deluxetable}

\clearpage

\appendix
\section{Additional SED Plots} \label{sec:appendix}

\begin{figure*}
\figurenum{A1}
\plotone{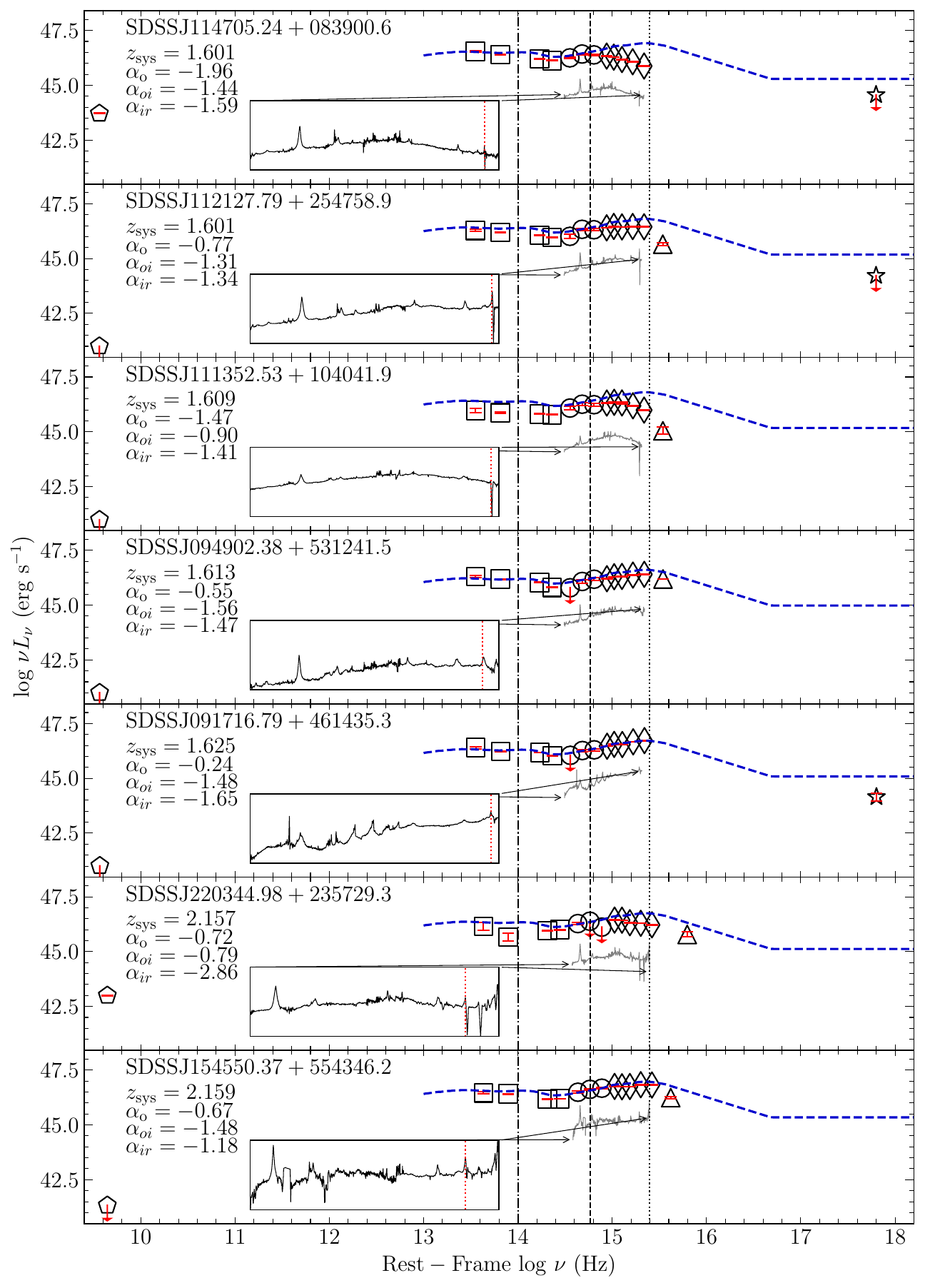}
\caption{SEDs for the remaining 58 GNIRS-DQS BAL quasars listed in Table \ref{tab:table1} in order of increasing $z_{\rm sys}$. Symbols represent: Radio = pentagons, WISE = squares, 2MASS = circles, SDSS = diamonds, GALEX = triangles, and X-ray = stars. The high-luminosity composite SED (dashed blue line) from K13 has been scaled to the flux density at 5100 \AA\ (log $\nu = 14.77$) in each source and over-plotted in each panel for reference. The gray spectrum in each panel, combined from M23 and \citet{Lyke20} data, is offset vertically for clarity; the inset panels zoom-in on these spectra with the \cfour~emission line marked by a red dotted line. The rest-frame wavelengths of 3 $\mu$m, 5100 \AA, and 1200 \AA\ are marked by dashed-dotted, dashed, and dotted lines, respectively.}
\end{figure*}
\begin{figure*}
\figurenum{A1}
\plotone{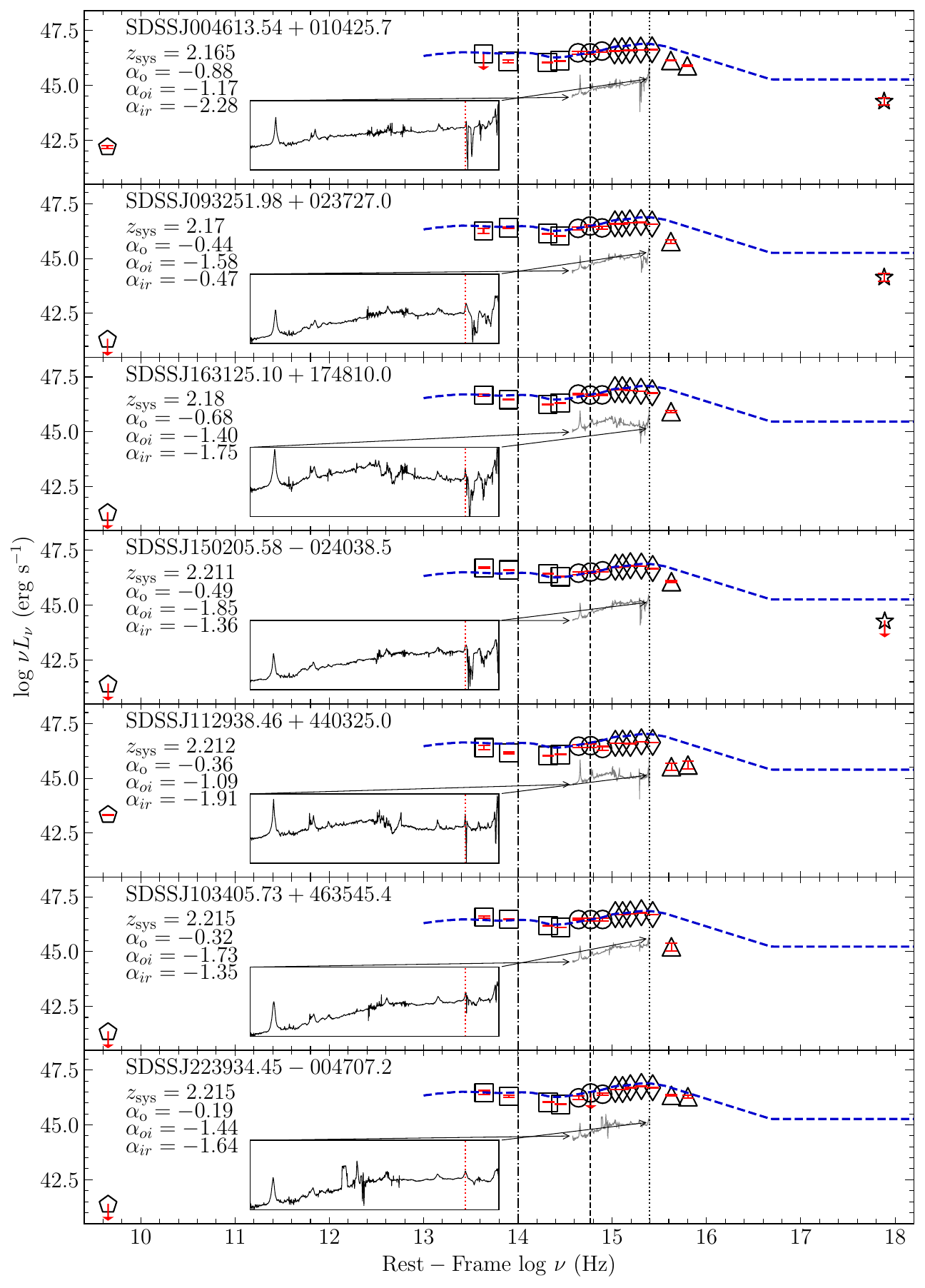}
\caption{cont.}
\end{figure*}
\begin{figure*}
\figurenum{A1}
\plotone{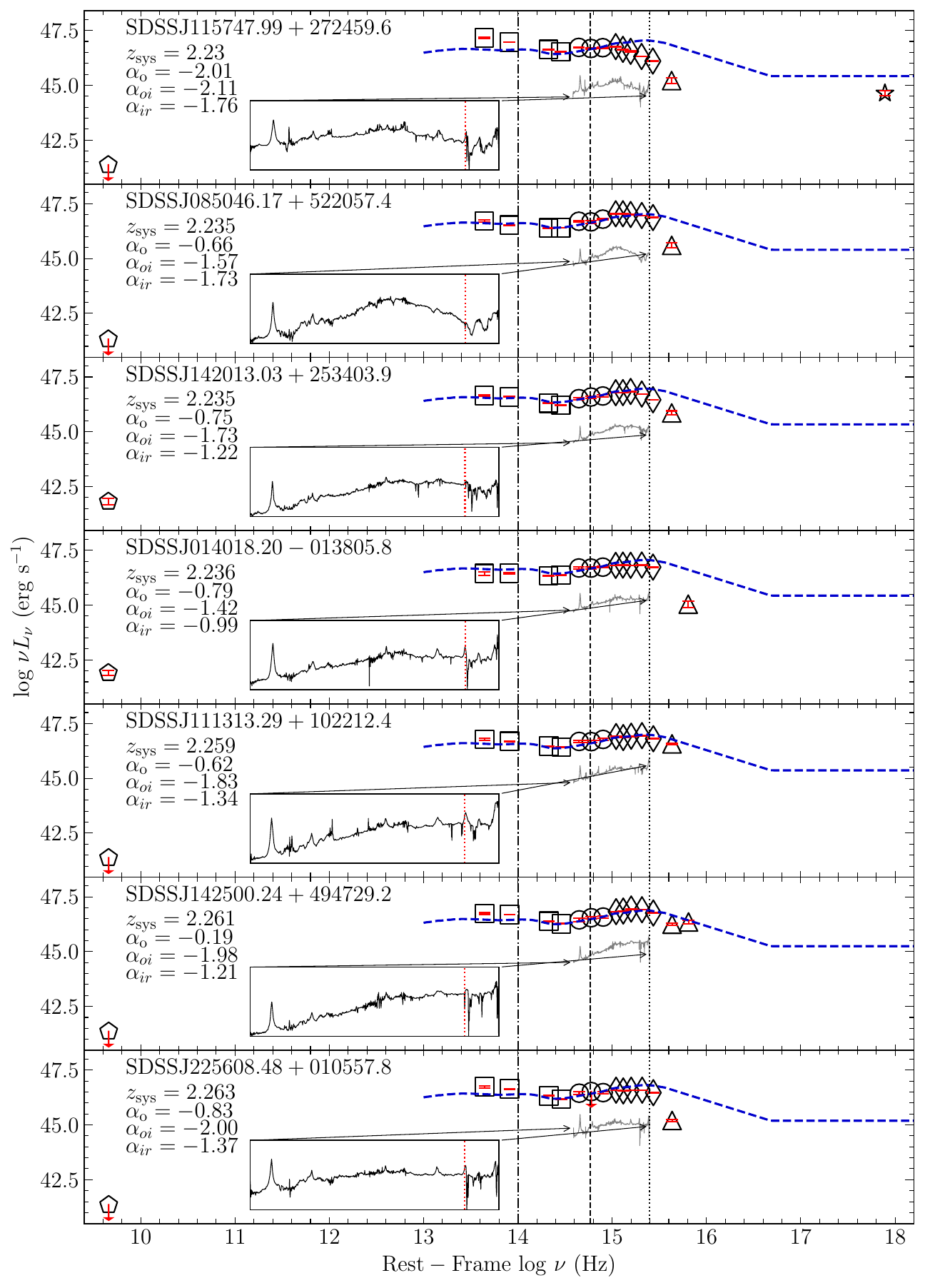}
\caption{cont.}
\end{figure*}
\begin{figure*}
\figurenum{A1}
\plotone{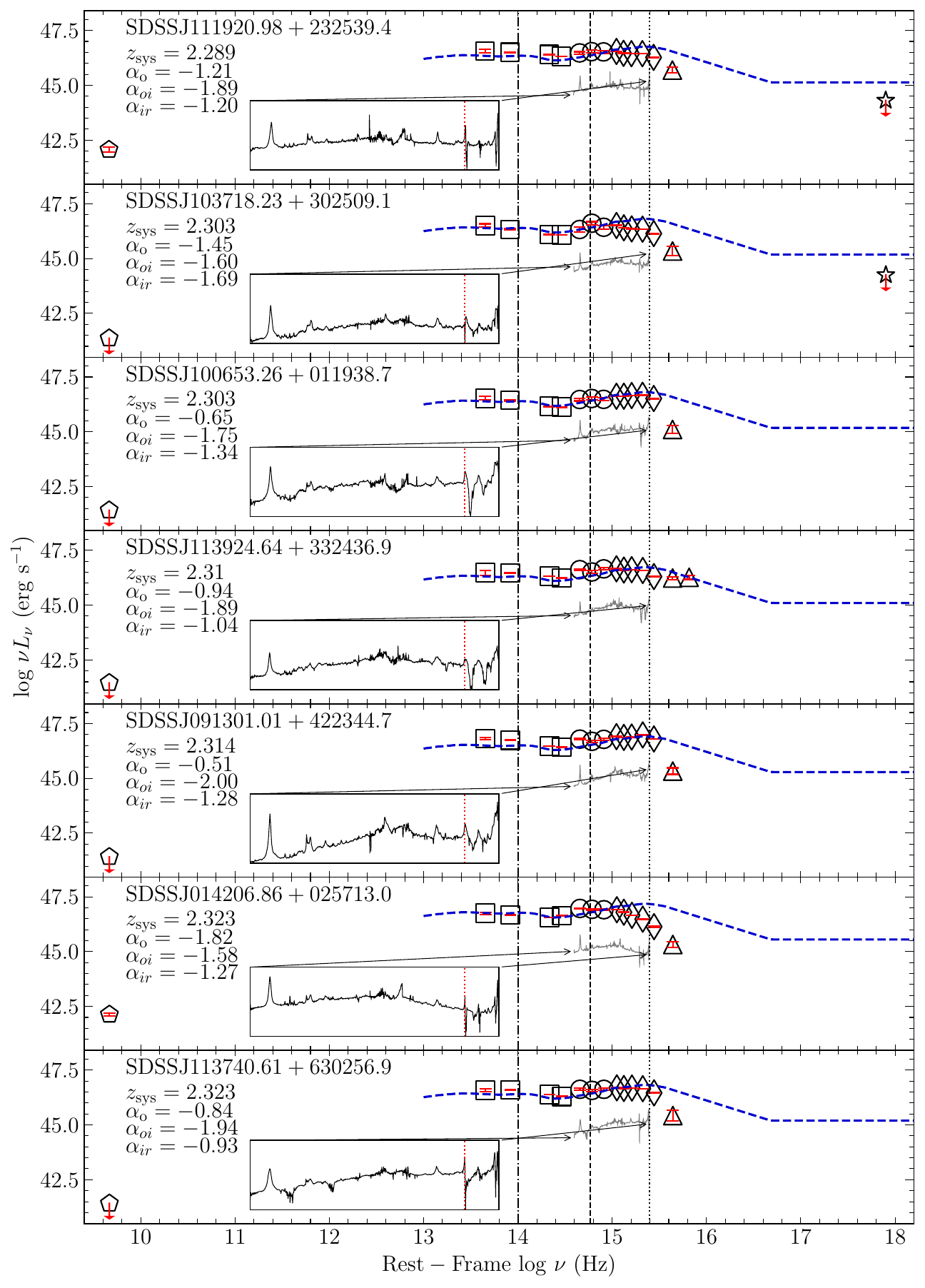}
\caption{cont.}
\end{figure*}
\begin{figure*}
\figurenum{A1}
\plotone{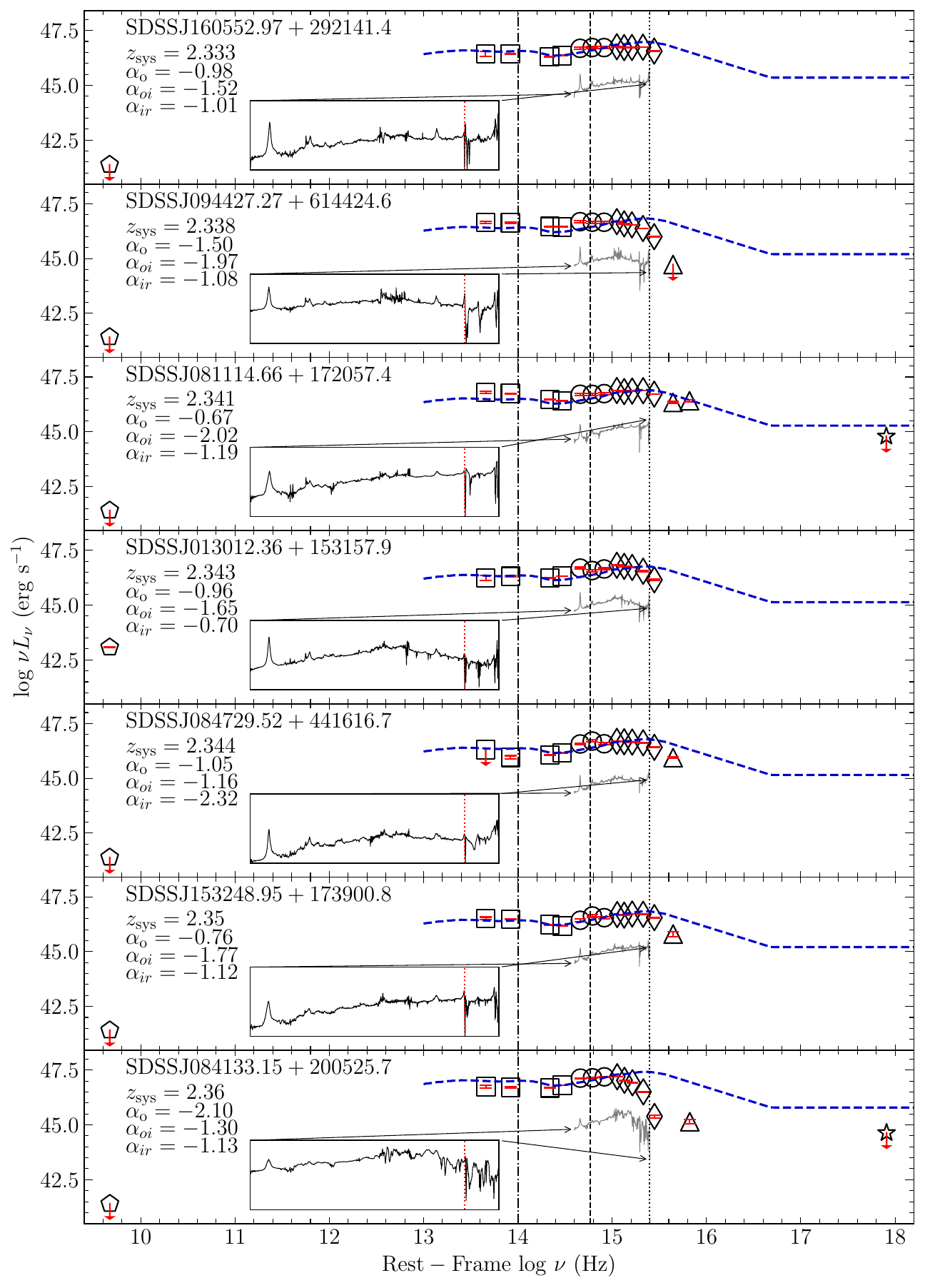}
\caption{cont.}
\end{figure*}
\begin{figure*}
\figurenum{A1}
\plotone{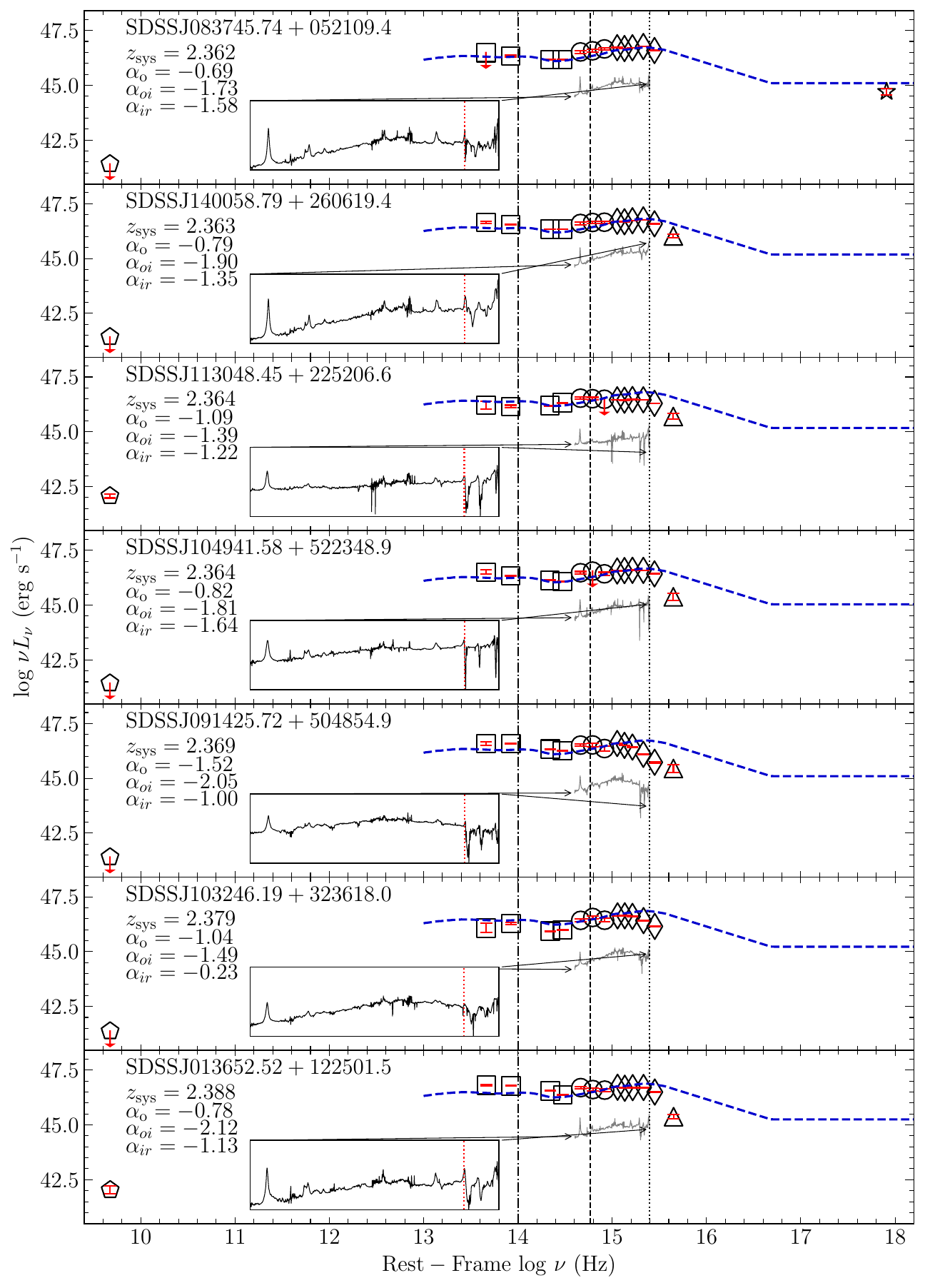}
\caption{cont.}
\end{figure*}
\begin{figure*}
\figurenum{A1}
\plotone{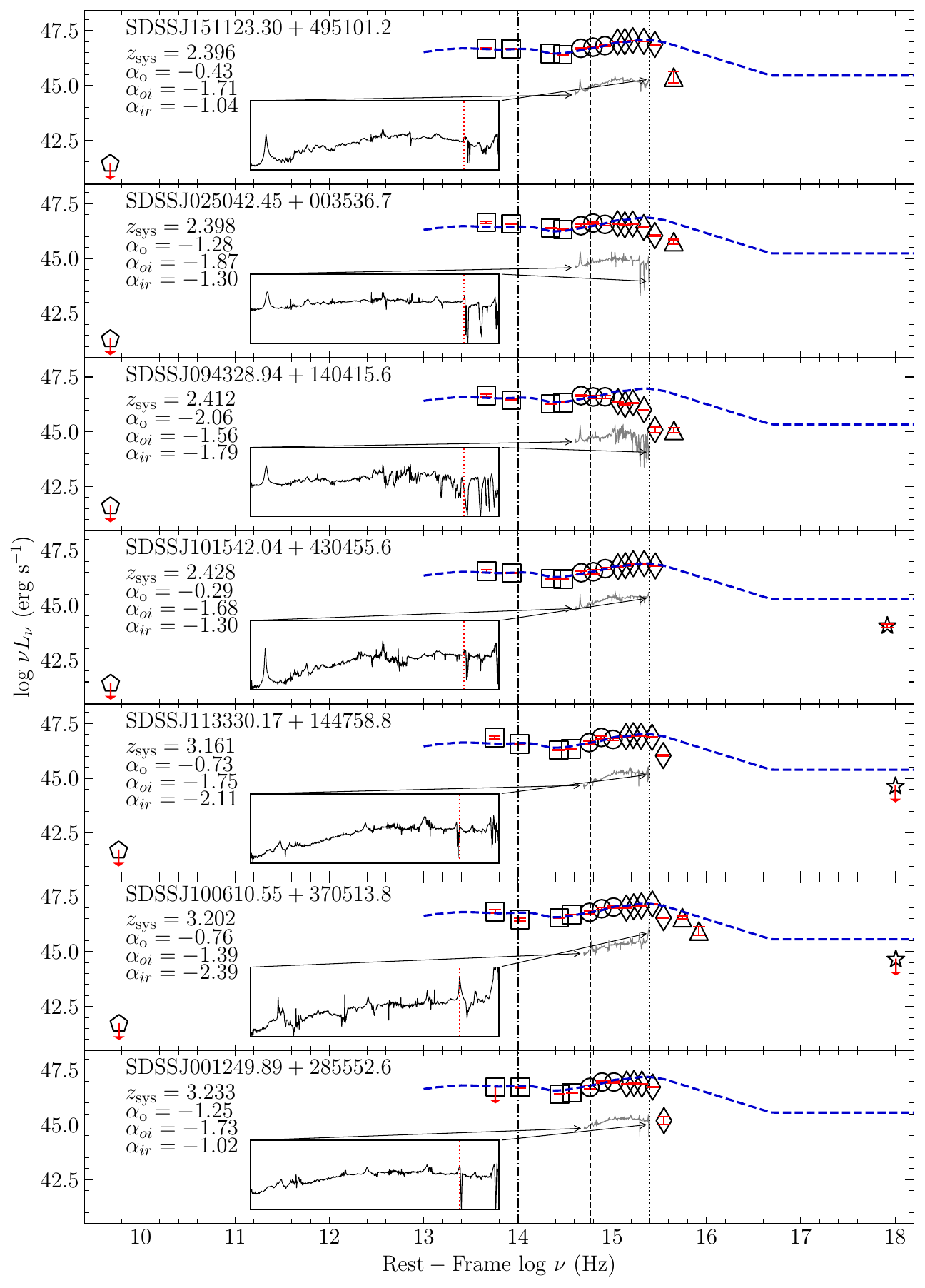}
\caption{cont.}
\end{figure*}
\begin{figure*}
\figurenum{A1}
\plotone{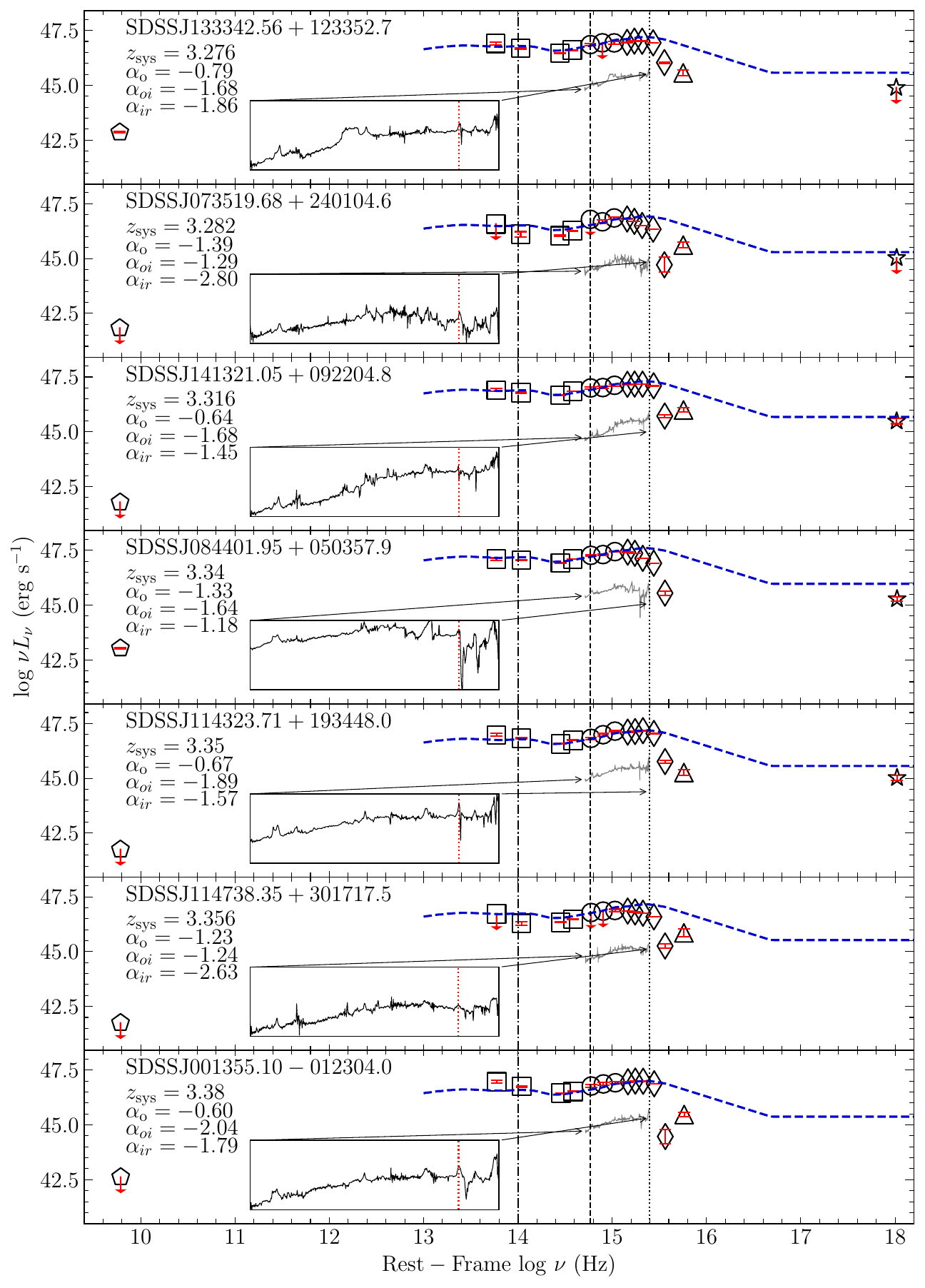}
\caption{cont.}
\end{figure*}
\begin{figure*}
\figurenum{A1}
\plotone{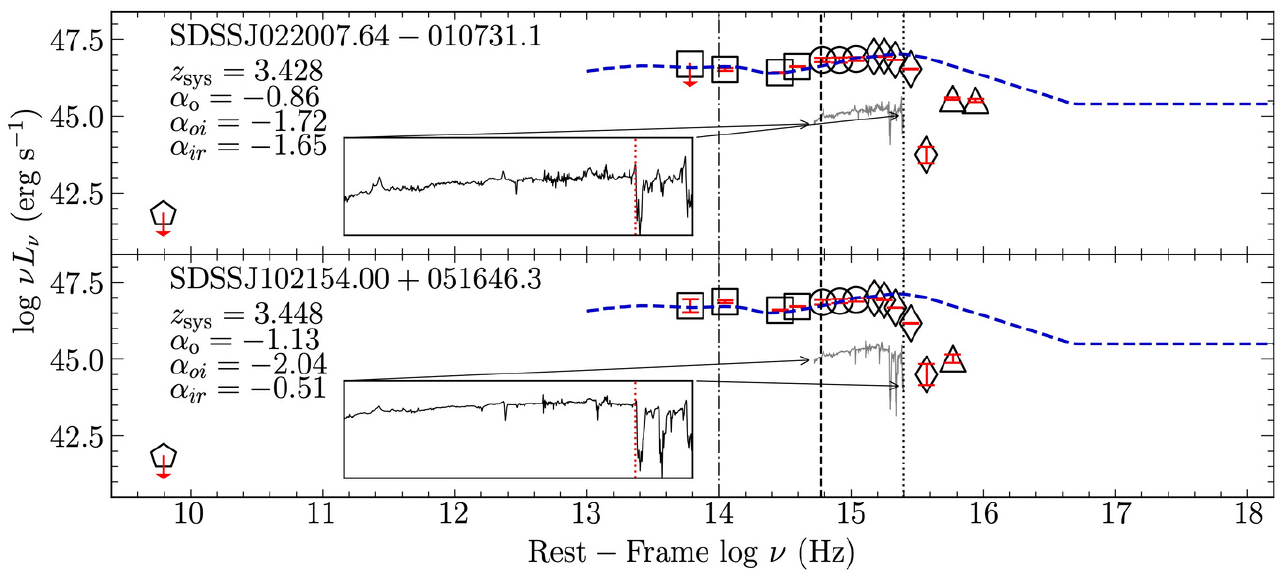}
\caption{cont.}
\end{figure*}

\end{document}